\documentclass[10pt,letterpaper]{IEEEtran}
\usepackage{color}
\usepackage{url}
\usepackage[latin9]{inputenc}
\usepackage{epstopdf}
\usepackage{amsthm}
\usepackage{amsmath}
\usepackage{amsthm}
\usepackage{amssymb}
\usepackage{graphicx}
\usepackage{subfigure}
\usepackage{wrapfig}
\usepackage{algorithmic}
\usepackage{algorithm}
\usepackage{mathrsfs}
\usepackage{float}
\usepackage{mathtools}
\usepackage{tabulary}
\usepackage{booktabs}
\usepackage{caption}
\graphicspath{{./}{./files/figures_eps/}}

\PassOptionsToPackage{bookmarks={false}}{hyperref}

\IEEEoverridecommandlockouts

\def\BibTeX{{\rm B\kern-.05em{\sc i\kern-.025em b}\kern-.08em
   T\kern-.1667em\lower.7ex\hbox{E}\kern-.125emX}}

\DeclareMathOperator{\vect}{vec}
\DeclareMathOperator{\rank}{rank}

\newcommand{\comment}[1]{ }

\newcommand\subparagraph{%
  \@startsection{subparagraph}{0}
  {\parindent}
  {0ex \@plus 0ex \@minus 0ex}
  {-1em}
  {\normalfont\normalsize\bfseries}}
\makeatother

\usepackage{titlesec}
\titlespacing*{\section}
{0pt}{1.2ex}{0ex} 
\titlespacing*{\subsection}
{0pt}{1ex}{0ex}  
\titlespacing*{\subsubsection}
{0pt}{0ex}{0ex} 

\begin{document}




\title{Leveraging Multiple Transmissions and Receptions for Channel-Agnostic Deep Learning-Based Network Device Classification}

\author{Nora Basha and Bechir Hamdaoui \\
\small School of Electrical Engineering and Computer Science, Oregon State University \\ \small \{bashano, hamdaoui\}@oregonstate.edu \\
}

\maketitle
\thispagestyle{plain}
\pagestyle{plain}

\begin{abstract}
The accurate identification of wireless devices is critical for enabling automated network access monitoring and authenticated data communication in large-scale networks; e.g., IoT. RF fingerprinting has emerged as a solution for device identification by leveraging the transmitter unique manufacturing impairments. Although deep learning is proven efficient in classifying devices based on the hardware impairments fingerprints, DL models perform poorly due to channel variations. That is, although training and testing neural networks using data generated during the same period achieve reliable classification, testing them on data generated at different times degrades the accuracy substantially, an already well recognized problem within the community. 
To the best of our knowledge, we are the first to propose to leverage MIMO capabilities to mitigate the channel effect and provide a channel-resilient device classification. We show that for AWGN channels, combining multiple received signals improves the testing accuracy by up to $30\%$.
We also show that for Rayleigh channels, blind channel estimation enabled by MIMO increases the testing accuracy by up to $40\%$ when the models are trained and tested over the same channel, and by up to $60\%$ when the models are tested on a channel that is different from that used for training.

\end{abstract}


\section{Introduction}
\label{sec:Introduction}
As the IoT paradigm is pervasively expanding into critical sectors including healthcare, home automation, and power grids, the flux of insecure devices connected to the Internet is widening the exposed attack surface of IoT networks \cite{singh_internet_2015, yeole_use_2016, rachedi_ieee_2016, noauthor_how_nodate}. Therefore, verifying the identity of the devices to prevent illegitimate devices from accessing and exploiting the network resources is becoming a necessity, and the need for a trustful device identification technique that is immune to spoofing and light enough to be implemented on the resources constrained IoT devices is becoming crucial \cite{zhang_iot_2014, chaabouni_network_2019}.
RF data-driven device fingerprinting has emerged as a promising technique for identifying and classifying devices using physical-level features that are much difficult, if not impossible, to spoof or replicate \cite{jian_deep_2020}. Transceiver hardware impairments that are inevitably inherited during device manufacturing provides a fingerprint for each device that, unlike high-layer features such as IP or MAC addresses \cite{smaini_rf_2012}, is immune to spoofing. Those hardware imperfections impair the transmitted waveforms in a way that provides transmitters with fingerprints and signatures that can uniquely separate them from one another \cite{sankhe_oracle_2019, zheng_fid_2019, nguyen_device_2011,elmaghbub2020widescan}.

Deep learning--or precisely Deep Neural Network (DNN)--based approaches have proven efficient in classifying devices from captured RF signals due to the DNNs' high dimensional mapping and ability to process data without the need of complicated features extraction. However, the strong assumption that the training and testing datasets are independent and identically distributed causes the DNN-based RF fingerprinting model accuracy to drastically degrade due to the time- and location-varying nature of the wireless channel that invalidates such a key assumption. For instance, the time-varying channel impulse response convoluted with the transmitted signal changes the signal and causes accuracy degradation for pre-trained DNN models \cite{al-shawabka_exposing_2020}. In addition, the channel overshadows the device fingerprints causing difficulties in training new models in the presence of fading channels. Efforts to design a channel-resilient RF fingerprinting techniques have taken two directions:
\begin{itemize}
    \item Communication system-oriented approaches where signal processing and communication approaches are applied at the transmitter or the receiver to remove the channel effect and enhance the deep learning model accuracy \cite{restuccia_deepradioid_2019,sankhe_oracle_2019}.
    \item Deep learning/data-driven approaches where data augmentation \cite{soltani_more_2020} or new neural network architecture designs \cite{ivanov_hybrid_2019} are introduced to mitigate the channel effect on RF fingerprinting.
\end{itemize}
In this paper, we propose a novel, channel-resilient device fingerprinting technique that tackles the wireless channel effect on DNN-based RF fingerprinting by leveraging the MIMO system capabilities in mitigating flat fading in Rayleigh channels and degraded SNR in AWGN channels. 
We first show that for AWGN channels, combining the multiple received noisy versions of the transmitted signal in a SIMO (transmitters each has a single antenna, receiver has multiple antennas) system improves the training accuracy substantially compared to the conventional SISO (all devices each has single antenna) system. We also show that when tested on a different AWGN channel with degraded SNR, the testing accuracy improves by up to $30\%$ compared to the SISO system. 
In addition, we show that combining the multiple received signals enabled via the SIMO system allows the DNN model to be trained on channels with lower SNR values without compromising the obtained classification accuracy, while at low SNR values, the same DNN architecture fails to distinguish between different devices as the impairments are overshadowed by the noisy channel effect in case of the conventional SISO approach. This improvement increases with the number of receiving antennas.
For flat fading channels, we show that Blind Channel Estimation performed by leveraging the MIMO capabilities combined with STBC (Space Time Block Coding) improves the training accuracy by $40\%$ over Rayleigh flat fading channel compared to SISO systems. We also show that when tested on a different flat fading channels, the accuracy of the DNN model is improved by up to $60\%$ compared to SISO systems. 

The rest of the paper is organized as follows. Section \ref{sec:related Works} presents previous related works. Section \ref{subsec:MIMO} offers a background about MIMO systems. Section \ref{subsec:AWGN} and Section \ref{subsec:Fading} explain the technique used for mitigating AWGN and flat fading channel models, respectively. Section \ref{sec:Eval} presents simulation results and evaluation. Finally, Section \ref{sec:conc} concludes the paper.

\section{Related Works}
\label{sec:related Works}
The inefficiency of deep learning-based device fingerprinting under time- and location-varying channel conditions has been a well recognized challenge within the device fingerprinting research community.
For instance, recent experimental results reported by Al-Shawabka \textit{et al.} \cite{ al-shawabka_exposing_2020} show that the wireless channel condition severely degrades the classification accuracy, dropping it from $85\%$ to $9\%$ using WiFi dataset and from $30\%$ to $17\%$ using Darpa dataset. They also show that equalizing the IQ data can improve the accuracy by up to $23\%$.
%
Sankhe \textit{et al.} \cite{sankhe_oracle_2019} benefit from the adaptivity feature of software defined radios and modify the transmitter chain of these radios such that their respective demodulated symbols acquire unique characteristics that make the CNN robust to channel changes (the signal unique characteristics dominate the channel changes). Restuccia \textit{et al.} \cite{restuccia_deepradioid_2019}, on the other hand, show that a carefully-optimized digital finite input response filter (FIR) at the transmitter's side, applying tiny modifications to the waveform to strengthen its fingerprint based on current channel conditions, can improve the accuracy from about $40\%$ to about $60\%$ in case of training on 5 devices. However, these works depend on modifying the transmitted signals by either adding artificial impairments that are immune to the channel variations, or by filtering to alter the transmitted signals to maximize the model accuracy, thereby resulting in a potential impact on the BER. In addition, these techniques require changes to be made at the transmitters' side.

Exploiting data augmentation techniques~\cite{soltani_more_2020} and exploring new CNN architectural designs~\cite{ivanov_hybrid_2019} have also been tried to mitigate the channel effect on RF fingerprinting. For instance, in~\cite{soltani_more_2020}, Slotani \textit{et al.} propose a data augmentation step within the training pipeline that exposes the DNN to many simulated channel and noise variations that are not present in the original dataset. Testing is later performed on the collected IQ traces. This data augmentation technique shows $75\%$ improvement in accuracy.

In this work, we propose a new framework that leverages MIMO systems hardware capabilities to mitigate the channel effect without the need for altering the transmitted signals, nor impacting the BER. To the best of our knowledge, we are the first to suggest leveraging MIMO capabilities to mitigate channel effect on RF fingerprinting. In this framework, we consider two channel models, AWGN and flat Rayleigh. For the AWGN channel model, we propose to combine the multiple received signals, and for the flat Rayleigh channel model, we propose to rely on blind partial channel estimation, enabled by MIMO and STBC, to improve the training and testing accuracy of the deep learning models, thereby increasing the device classification performance substantially.



\section{Background on MIMO and Spatial Diversity}
\label{subsec:MIMO}
Among other well known benefits, MIMO (Multiple-Input Multiple-Output) links improve SNR (signal-to-noise ratio) of multipath fading channels through spatial diversity by means of combining the output signals received on multiple uncorrelated antenna elements in presence of fading caused by multipath propagation \cite{winters_diversity_1998,clerckx_mimo_2013}.  
The improvement in the SNR achieved through diversity is characterized by~\cite{clerckx_mimo_2013}:
\begin{itemize}
    \item Array gain, which measures the increase in average output SNR relative to the single-branch average SNR.
    \item Diversity gain, which measures the increase in the error rate slope as a function of the SNR. 
\end{itemize}
In other words, array gain increases the average SNR, whereas diversity gain makes the probability density function of the instantaneous SNR more concentrated around its average~\cite{zaidi_chapter_2018}.
Two types of diversity could be realized via MIMO: receive diversity and transmit diversity.

\subsubsection{Receive Diversity}
This can be realized with SIMO (single-input, multiple-output) links, i.e., when the receiver is equipped with multiple antennas and the transmitter is equipped with a single antenna. Receive diversity could be obtained by two combining methods: (i) selection combining, where the receiving antenna element whose signal offers the highest SNR is chosen for detection and (ii) gain combining, where all receiving antenna signals are optimally combined to increase the overall SNR~\cite{clerckx_mimo_2013}. Compared to conventional single receiving antenna systems, receive diversity yields a higher average SNR value (array gain), and a lower symbol error rate.
In our studied device identification problem, the SNR improvement due to spatial diversity is exploited for  mitigating the degradation of the classification accuracy that occurs due to AWGN channel variations (i.e., the DNN model is tested on a channel whose conditions are different from those used for training). Therefore, MIMO is leveraged in our framework to devise classification approaches that are agnostic to channel condition variations. 

\subsubsection{Transmit Diversity and Space-Time Block Coding}
Transmit diversity is enabled by the multiple transmit antenna capability of MIMO links, along with pre-processing or pre-coding capability. 
One main difference between receive and transmit diversity is that the channel state needed for performing the diversity technique can be easily estimated at the receiver side, but not so at the transmitter side, making the transmit diversity technique more challenging to perform~\cite{clerckx_mimo_2013}. 
%
Therefore, schemes that achieve transmit diversity have taken two directions. (i) When there is feedback between the receiver and the transmitter and the transmitter has a perfect channel knowledge, beamforming is used at the transmitter to optimize the SNR and achieve both array and diversity gains. This is performed by multiplying the transmitted signal vector by a weight vector before transmitting, where the weight vector is obtained by maximizing the SNR given perfect channel knowledge at the transmitter~\cite{godara_application_1997}.
(ii) When no knowledge is available at the transmitter about the channel, STBCs are used to achieve diversity gain~\cite{clerckx_mimo_2013}.
STBCs have been established as an efficient method to enhance wireless communication systems performance.
In essence, the STBC technique consists of spreading information symbols in space using multiple transmitting antennas and in time using pre-coding to achieve transmit diversity \cite{tarokh_space-time_1999, foschini_layered_1996, winters_diversity_1998, clerckx_mimo_2013}. One of the earliest transmit diversity schemes is the Alamouti scheme for two transmit antennas~\cite{alamouti_simple_1998} and thus with time diversity of two.
When the receiver is only equipped with one single antenna, Alamouti scheme does not achieve an array gain as no channel knowledge is available. However, for i.i.d. Rayleigh channels, a diversity gain is achieved and the symbol error rate is reduced. To achieve array gain, the receiver must be equipped with more than one antenna.

STBCs make it possible to apply blind estimation of the channel by observing only the received signals \cite{swindlehurst_blind_2002, ammar_blind_2007, peken_blind_2017}, and leveraging this MIMO's blind estimation capability with STBC is what we propose in this paper for improving the resiliency of deep learning based RF fingerprinting to channel condition variations.

\section{Leveraging MIMO for Channel-Agnostic Wireless Device Identification}
\label{proposed}
As mentioned in the introduction, prior deep learning based fingerprinting approaches suffer from the impact of time- and location-varying channel conditions. In other works, although they show promising results when the learning models are trained and tested on data collected under the same channel conditions, these approaches perform poorly when training and testing are done on data collected under different channel conditions. 
In this section, we propose new fingerprinting approaches that are resilient to channel condition changes. The novelty of our techniques lies in leveraging the capabilities offered by MIMO systems to mitigate the distortions in the received RF signals caused by the wireless channel, thereby making the learning models agnostic to the underlying channel. 

We consider both of the AWGN and flat fading Rayleigh channel models in this framework, which we present next.


\subsection{AWGN Channel-Agnostic Device Fingerprinting}
\label{subsec:AWGN}
We begin this section by considering an AWGN channel model. The more realistic fading channel model is considered in the next section. For AWGN channels, we leverage the capability of SIMO systems (the receiver is equipped with multiple antennas but transmitters are each single-antenna equipped) to combat the impact of channel variations. For a SIMO system, with $L$ receiving antennas, sending a signal $s(t)$ over an AWGN channel with zero mean and $\sigma^2$ noise power, the received signal by the i$^{th}$ receiving antenna is
$r_i(t)=s(t)+n_i(t)$, where $n_i(t)$ is the noise seen at antenna $i$.
Our technique proposed for AWGN channels consists of averaging the received signals over all the receiving antennas to achieve SNR gains that mitigate the channel noise, and using this averaged received signal $r(t)=s(t)+ \frac{1}{L}  \sum_{i=1}^{L} n_{i}(t)$ for training the DNN models. This yields a classification accuracy that is less sensitive to noise, and more agnostic to AWGN channels. In addition, testing the models over an AWGN channel with a lower SNR using the averaged $r(t)$ is less affected by the noise, and the reduction in the testing accuracy due to the change of the channel is lesser than what a single-antenna receiver achieves. 
%
%
%
Since the noise samples received by the antenna elements are i.i.d., the noise power decreases by a factor of $L$ and $r(t)$ is considered an unbiased, consistent estimator of the originally transmitted waveform $s(t)$. Increasing the number of receiving antennas $L$ makes $r(t)$ more immune to the channel noise when compared to a waveform received by a single antenna.




\subsection{Flat Fading Channel-Agnostic Device Fingerprinting}
\label{subsec:Fading}
For Rayleigh fading channels, we exploit combined MIMO and STBC capabilities to improve the training and testing accuracy of the deep learning models by mitigating the RF data distortions caused by the Rayleigh channel, as we explain next. 
When a MIMO system transmits an STBC signal over a flat fading channel, each receiving antenna receives a signal that is a mixture of the signals transmitted by all the transmitting antennas, and each of the transmitted signal contributes to the mixture with a weight dictated by the channel matrix. The problem of estimating the transmitted signals given the received signals and the properties of the transmitted signals, referred to as the blind source separation/blind channel estimation problem, essentially boils down to finding the inverse of the channel matrix, which can then be used to recover the transmitted signals~\cite{hyvarinen_independent_2001, ammar_blind_2002,ammar_blind_2007}.

\subsubsection{The Channel Matrix Estimation}
Consider a MIMO link, with $M$ transmitting and $L$ receiving antennas, sending an STBC signal over a flat fading channel defined by the $L \times M$ channel matrix $\mathbf{H}$. STBC essentially maps a data vector $\mathbf{s}$ with $n$ symbol entries into an $M \times K \quad (K\leq M)$ code matrix $\mathcal{C}$, where $K$ is the time diversity of the code. Each column of $\mathcal{C}$ is a linear mapping of the real part $\mathbf{s}_R$ and the imaginary part $\mathbf{s}_I$ of the vector $\mathbf{s}$. In other words, each of the $K$ columns in $\mathcal{C}$ is given by:
\[\underbrace{[\mathbf{A}_k \quad  j\mathbf{B}_k]}_{\triangleq \mathbf{C}_k} \underbrace{
\begin{bmatrix}
\mathbf{s}_R  \\
\mathbf{s}_I
\end{bmatrix}}_{\triangleq \mathbf{x}}, \quad  k=1,2,...,K.\]
where $\mathbf{x}$ is a $2n$ vector, $\mathbf{A}_k$ and $\mathbf{B}_k$ are $M \times n$ matrices derived from the STBC and known to the transmitter and receiver~\cite{swindlehurst_blind_2002,ammar_blind_2007,tarokh_space-time_1999}, and
$\otimes$ denotes the Kronecker product.
For instance, for the Alamouti coding scheme~\cite{alamouti_simple_1998},  
\begin{equation*}
\small
    \begin{split}
    \mathcal{C}= \begin{bmatrix}
    c_{1} & -c_{2}^* \\
    c_{2} & c_{1}^*
    \end{bmatrix}\!\!,\;
    \mathbf{A}_{1} = \mathbf{B}_1 = \begin{bmatrix}
    1 & 0\\
    0 & 1\\
    \end{bmatrix}\!\!,\;  
    \mathbf{A}_{2} = -\mathbf{B}_2 = \begin{bmatrix}
     0 &  -1 \\
    1 & 0 \\
    \end{bmatrix}
    \end{split}
\end{equation*}

The $K$ columns of $\mathcal{C}$ are transmitted in $K$ time epochs by the transmitting antennas.
Assuming that the channel is static over the duration of $K$ epochs, the received signal vector is
\begin{equation} \nonumber 
\mathbf{r}= \begin{bmatrix}
\mathbf{HC}_1  \\
\mathbf{HC}_2 \\
\vdots \\
\mathbf{HC}_K
\end{bmatrix} \mathbf{x} + \mathbf{n}= (\mathbf{I}_K \otimes \mathbf{H})\underbrace{\begin{bmatrix}
\mathbf{C}_1  \\
\mathbf{C}_2   \\
\vdots \\
\mathbf{C}_K
\end{bmatrix}}_{\triangleq \mathbf{C}} \mathbf{x}+ \mathbf{n}
\end{equation}
where $\mathbf{n}$ is a $2n$ noise vector, $\mathbf{I}_k$ is $k \times k $ identity matrix, $\mathbf{HC}_i$ is an $L \times 2n$ matrix, and $\mathbf{C}$ is an $MK \times 2n$ matrix. If the channel is also static over $l$ received data symbol vectors $\mathbf{r}_i$ each with $K$ epochs, the obtained received matrix becomes
\begin{equation} \label{eq2}
    \mathbf{R} = (\mathbf{I}_K \otimes \mathbf{H})\mathbf{CS} + \mathbf{N}
\end{equation}
where $\mathbf{R}\triangleq \begin{bmatrix}
        \mathbf{r}_1 & \mathbf{r}_2 & \hdots & \mathbf{r}_l
        \end{bmatrix}$ 
is a $KL \times l$ received matrix and 
$\mathbf{S}\triangleq  \begin{bmatrix}
        \mathbf{x}_1 & \mathbf{x}_2 & \hdots & \mathbf{x}_l \end{bmatrix}$ 
is a $2n \times l$ data symbol matrix.
%
%
The $l$ subscript indicates the block transmission on $l$ symbols, and the goal is to find the $L \times M$ channel matrix $\mathbf{H}$.
If $\mathbf{N}_L$ is the left null space of $\mathbf{R}$, then
\begin{equation} \label{eq3}
    \mathbf{N}^H_L(\mathbf{I}_K \otimes \mathbf{H})\mathbf{CS} = \mathbf{0}
\end{equation}
where $\{.\}^H$ denotes the conjugate transpose operation.
Assuming $\mathbf{S}$ is full raw rank, i.e. data is persistently exciting, Eq.~\eqref{eq3} reduces to
\begin{equation} \label{eq4}
     \mathbf{N}^H_L(\mathbf{I}_K \otimes \mathbf{H})\mathbf{C}= \mathbf{0}
\end{equation}
Eq.~\eqref{eq4}, also called the blind equation, represents a homogeneous linear equation in the unknown $\mathbf{H}$, and the uniqueness of the solution depends on the matrix $\mathbf{C}$ of STBC and the rank of the matrix $\mathbf{H}$~\cite{ammar_blind_2002}.

Our proposed technique relies on the blind estimation algorithm proposed in~\cite{ammar_blind_2007} to estimate the channel matrix, which starts by reshaping Eq.~\eqref{eq4}, using the vector property of the Kronecker product, as
\begin{equation} \label{eq5}
    \mathbf{C}^T \otimes \mathbf{N}^H_L  \cdot \vect(\mathbf{I}_K\otimes \mathbf{H})=0
\end{equation}
where $\{.\}^T$ denotes the transpose operation and $\vect(.)$ is the matrix vectorization operation.
To decouple the vector form of the unknown $\mathbf{H}$, we first write, 
for $k =1,\hdots,K$,
\begin{equation} \label{eq6} \small
\vect(\mathbf{I}_K \otimes \mathbf{H}) \! = \!\! \begin{bmatrix}
                    \mathbf{I}_M \otimes \mathbf{E}_1 \\
                    \mathbf{I}_M \otimes \mathbf{E}_2 \\
                    \vdots \\
                    \mathbf{I}_M \otimes \mathbf{E}_K \\
                    \end{bmatrix} \!\! \mathbf{h} 
                    \mbox{\; with }
\mathbf{E}_k \! \triangleq \!\! \begin{bmatrix}
                      \mathbf{0}_{L(k-1)\times L} \\
                           \hline\\
                            \mathbf{I}_L \\
                            \hline \\
                          \mathbf{0}_{L(K-k)\times L} 
                           \end{bmatrix}
\end{equation}
\comment{
\begin{multline} \label{eq6}
\vect(\mathbf{I}_K \otimes \mathbf{H})  =\begin{bmatrix}
                    \mathbf{I}_M \otimes \mathbf{E}_1 \\
                    \mathbf{I}_M \otimes \mathbf{E}_2 \\
                    \vdots \\
                    \mathbf{I}_M \otimes \mathbf{E}_K \\
                    \end{bmatrix} \mathbf{h}\\
\mathbf{E}_k  \triangleq \begin{bmatrix}
                      \mathbf{0}_{L(k-1)\times L} \\
                           \hline\\
                            \mathbf{I}_L \\
                            \hline \\
                          \mathbf{0}_{L(K-k)\times L} 
                           \end{bmatrix}, \quad K =1,\hdots,K
\end{multline}
}
and then substitute Eq.~\eqref{eq6} into Eq.~\eqref{eq5} to get
\begin{equation} \label{eq7}
   \overbrace{(\sum_{k=1}^{K} \mathbf{C}^T_k \otimes (\mathbf{N}^H_L \mathbf{E}_k))}^ {\triangleq \mathbf{\Delta}} \mathbf{h}= 0.
\end{equation}
Eq.~\eqref{eq7} shows that $\mathbf{h}$ lies in the right null space of $\mathbf{\Delta}$, and when the channel is uniquely identifiable up to a single complex scalar ambiguity, $\mathbf{h}$ spans the null space of $\mathbf{\Delta}$.
Singular value decomposition can be used to find the basis of the null space of $\mathbf{\Delta}$. The right singular vector corresponding to the smallest singular value of $\mathbf{\Delta}$ is the estimated channel vector $\mathbf{\hat h}$ with $\mathbf{\hat h}= \alpha \mathbf{h}$ for some $\alpha \in \mathbb{C}$.
%
This algorithm can estimate the channel matrix $\mathbf{H}$ up to a single complex ambiguity provided that Eq.~\eqref{eq4} has a unique solution~\cite{ammar_blind_2007}, and 
%
the channel is uniquely identifiable up to a single complex scalar ambiguity if and only if $\rho=1$ when either $\mathbf{H}$ is full rank, or $\mathbf{H}$ is column rank deficient but $\rank(\mathbf{H})> M-r_{min}$~\cite{ammar_blind_2007}.
Here $\rho$ and $r_{min}$ depend only on the used STBC and are shown in Table~\ref{Tb1} for some generic STBCs~\cite{ammar_blind_2007}.
Given that $\rank(\mathbf{H})= \min(M,L)$, these conditions on the channel rank constrain the minimum number of receiving antennas required to estimate the channel blindly such that $L \geq M-r_{min}+1$.
For instance, using Tarokh STBC $\mathcal{O}_3$ (see Table~\ref{Tb1}) to transmit QPSK symbols requires at least 3 receiving antennas for the channel to be blindly estimated up to a single complex scalar ambiguity.

\comment{
If $\mathbf{R}$ in Eq.~\eqref{eq4} has a nontrivial left null space and  
$\mathbf{\Tilde{C}} \triangleq \begin{bmatrix}
\mathbf{C}_1 & \mathbf{C}_2 & \hdots & \mathbf{C}_K
\end{bmatrix}$
has a full row rank, then the channel is uniquely identifiable up to a single complex scalar ambiguity when the following conditions are satisfied~\cite{ammar_blind_2007}:
\begin{itemize}
\item For full rank channel matrix $\mathbf{H}$, the channel is uniquely identifiable up to a complex scalar ambiguity if and only if $\rho=1$.
\item For column rank deficient channel matrix $\mathbf{H}$, if $rank(\mathbf{H})> M-r_{min}$, the channel is uniquely identifiable up to a complex scalar ambiguity if and only if $\rho=1$.
\end{itemize}
}

\begin{table}\scriptsize
\centering 
\caption{\small Summary of $\rho$ and $r_{min}$ for some generic STBCs~\cite{ammar_blind_2007}} 
\begin{tabular}{|c|c|c|c|c|c|} 
\hline 
STBC & $(M,n,K)$ & $\rho$ & $r_\text{min}$ & $r_\text{min}^\prime$ & rank(C)\\
\hline 
$\mathcal{O}_1$ (Alamouti) & $(2,2,2)$ & 4 & 1 & 2 & 4\\ 
\hline
$\mathcal{O}_2$ (Tarokh) & $(3,3,4)$ & 1 & 3 & 3 & 6\\
\hline
$\mathcal{O}_3$ (Tarokh) & $(3,4,8)$ & 1 & 1 & 3 & 8\\ 
\hline 
\end{tabular}
\label{Tb1}
\end{table}
In practice, the received signal matrix $\mathbf{R}$ is affected by additive white Gaussian noise, so it is required to separate the noise and the signal subspaces before using the blind algorithm~\cite{ammar_blind_2002} using the subspace approach~\cite{moulines_subspace_1995}.
To solve the remaining ambiguity and completely estimate the channel, additional information about the channel matrix is obtained by sending pilots known for the receiver. The pilots assist the blind algorithm to completely identify the channel~\cite{ammar_blind_2007}. In our proposed RF/device classification technique, we apply the blind partial estimation up to a single complex ambiguity without using pilots, and then exploit deep learning generalizing abilities to classify different wireless devices using received RF signal data.

\subsubsection{The Proposed MIMO-Enabled Device Fingerprinting}
The proposed MIMO-enabled technique for Rayleigh fading channels uses the blind partial channel estimation enabled by the MIMO system proposed in~\cite{ammar_blind_2007} to estimate the channel matrix. We set the MIMO hardware and the STBC at the minimum requirements to guarantee a blind channel estimation from the received signals up to a single complex ambiguity, i.e. $\rho=1$. Once the channel is blindly estimated up to a single complex ambiguity, the resolved signals are sampled and used for training and testing the deep learning model used for classification. In our proposed technique, we use CNN to learn from training data and then generalize on unseen data to overcome the remaining ambiguity and capture the impairments effect on the reconstructed waveforms.


\section{Performance Evaluation and Analysis}
\label{sec:Eval}
For both fingerprinting techniques proposed for mitigating the impact of AWGN and flat fading channels on classification accuracy, we use MATLAB R2020b to build our wireless communication model and generate the IQ datasets for training, validation, and testing from IEEE 802.11n waveforms. The IQ data were collected form 10 simulated RF devices uniquely impaired with the impairments set in Table~\ref{Tb2}. The impairments are set slightly different to mimic the slight differences among devices. For each device, we collected $5000$ frames, with each frame having the size of $160$. Then we split the real and the imaginary parts of the signal and reshaped the frames as $2 \times 160$ vectors to be fed to the input layer of the CNN. The dataset was divided into $80\%$ for training, $10\%$ for validation and  $10\%$ for testing.  
\begin{table*}[!t]
\centering 
\caption{\small Hardware impairments used to simulate 10 different devices} 
\resizebox{\textwidth}{!}{\begin{tabular}{|c|c|c|c|c|c|c|c|c|} 
\hline 
Device & Phase Noise & Frequency Offset & IQ Gain Imbalance & IQ Phase Imbalance & AMAM & AMPM & Real DC Offset & Imaginary DC Offset\\
\hline 
DV1& -60 & 20 & 0.08 & 0.1 & [2.1587,1.1517] & [4.0033,9.104] & 0.1 & 0.15 \\
\hline
DV2& -60.15 & 20.01 & 0.1 & 0.09 & [2.1687,1.1617] & [4.1033,9.124]& 0.11 & 0.14 \\
\hline
DV3& -59.9& 20.2& 0.09 &0.09  & [2.1789,1.1317] & [4.0933,9.151] & 0.1 & 0.11 \\
\hline
DV4& -60.1 & 20  & 0.108 & 0.109 & [2.1987,1.1217] & [4.1033,9.194]  & 0.1 & 0.1 \\
\hline
DV5& -60  & 20.09& 0.1 & 0 & [2.1587,1.1717] & [4.093,9.094] & 0.089 & 0.1008\\
\hline
DV6& -59.95 & 20.1 & 0.12 & 0.15 & [2.1487,1.1117] & [4.1033,9.156] & 0.1 & 0.098  \\
\hline
DV7& -59.93&  20.11& 0.11 & 0.11 & [2.1897,1.1237] & [4.1133,9.135] & 0.111 & 0.1011 \\
\hline
DV8& -60.13 & 20.099 & 0.101 & 0.14 & [2.1387,1.1627] & [4.1533,9.096] & 0.12  & 0.099
 \\
\hline
DV9&-59.89  & 19.9& 0.099 & 0.08 & [2.1548,1.1917] & [4.09833,9.10056] & 0.09 & 0.0999 \\
\hline
DV10& -59.91 & 19.98  & 0.111 & 0.105 & [2.1777,1.09874] & [4.0987,9.123] & 0.101 & 0.10015 \\
\hline 
\end{tabular}}
\label{Tb2}
\end{table*}

\subsection{CNN Architecture}
We used the CNN architecture in~\cite{sankhe_oracle_2019}
as a benchmark to evaluate the proposed MIMO-enabled approaches and compare them to the SISO/conventional approach. 
The CNN architecture consists of two convolution layers and two fully connected layers. The $2 \times 160$ input is fed into the first convolution layer that consists of $50$ $1 \times 7$ filters. This layer produces $50$ features maps from the entire input. The second convolution layer consists of $50$ $2 \times 7$ filters, where each filter is convoluted with the 50-D volumes obtained from the first layer. The second convolution layer learns variations over both  I and Q dimensions. Each convolution layer is followed by a ReLU activation function to add non-linearity, and a 2-stride max pooling layer to prevent overfitting. The first fully connected layer consists of $256$ nodes whose output are fed into the second fully connected layer of $80$ nodes. Each fully connected layer has a ReLU activation.
The last layer is a soft max classifier to generate the classification probabilities. At the classifier output, the cross-entropy loss is calculated and the back-propagation algorithm is used to find the network parameters that minimize the prediction error. The CNN uses Adam optimizer with a learning rate of $0.0001$.
\comment{
\begin{figure}
\centering
\includegraphics[width=.75\columnwidth]{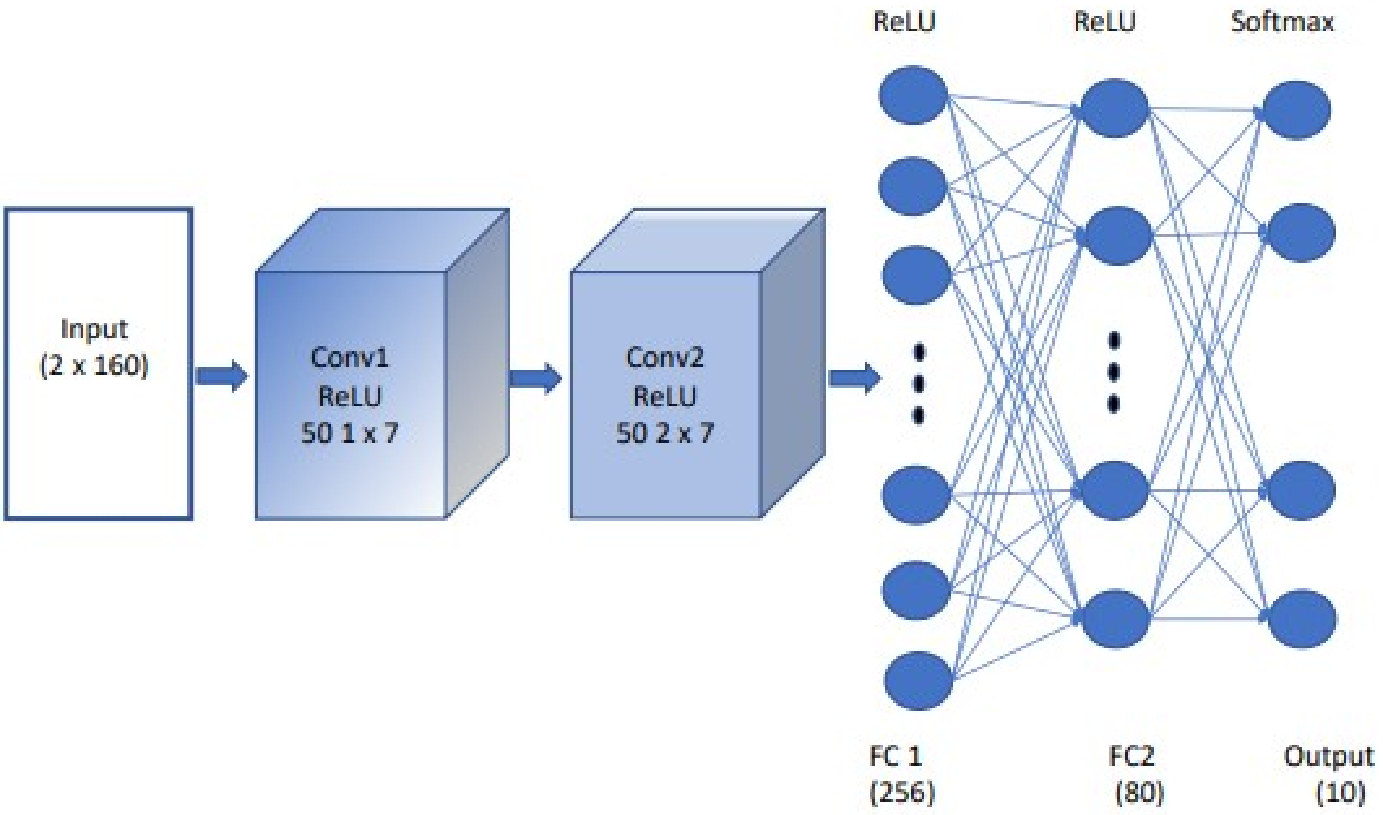}
\caption{CNN Architecture}
\label{CNN Architecture}
\end{figure}
}

\subsection{Performance Metrics}
\label{subsec:metrics}
We consider the following metrics in this evaluation.
\begin{itemize}
    \item \textbf{Training Accuracy}, the percentage of the correctly classified samples to the total number of samples in the training datatset.
    \item \textbf{Testing Accuracy}, the percentage of the correctly classified samples to the total number of samples in the testing datatset.
    \item \textbf{Relative Different channel Testing Gap (RDTG)}, the percentage of reduction occurred in the testing accuracy when the testing channel is different from the training channel. Precisely, RDTG is defined as
\begin{equation*}
\resizebox{0.44\textwidth}{!}{$ \textrm{RDTG} = \frac{\textrm{same channel testing acc.} - \textrm{different channel testing acc.} }{\textrm{same channel testing acc.}} \%$}
\end{equation*}
If the classification technique is perfectly channel-agnostic, then RDTG is zero, and the deviation from zero quantifies the effect of the channel variation on accuracy. RDTG values are only calculated when training and testing channels are different.
\end{itemize}
In the evaluation of the proposed fading channel-agnostic technique, we also vary the following parameters:
\begin{itemize}
    \item \textbf{Training APG}, the average path gain (APG) of the Rayleigh flat fading channel used for training. APG is varied from -20 dB to 20 dB.
    \item \textbf{Testing APG}, the average path gain (APG) of the Rayleigh flat fading channel used for testing. APG is varied from -20 dB to 20 dB.
    \item \textbf{Training MDS}, the maximum Doppler shift (MDS) of the Rayleigh flat fading channel used for training. MDS is varied from 1/2000 Hz to 1 Hz.
    \item \textbf{Testing MDS}, the maximum Doppler shift (MDS) of the Rayleigh flat fading channel used for testing. MDS is varied from 1/2000 Hz to 1 Hz.
\end{itemize}

In this section, we study the resiliency of the proposed MIMO-enabled classification approaches to wireless channel variations and compare them against the conventional SISO fingerprinting technique. Here again, as widely adopted, we use MIMO to refer to the case when the receiver and all transmitters are each equipped with multiple antennas, SIMO to refer to the case when the receiver is equipped with multiple antennas and the transmitters being classified are each equipped with a single antenna, and SISO to refer to the conventional case when the receiver and all transmitters are each equipped with a single antenna.

\subsection{AWGN Channel Results}
In this section, we consider studying the benefit of exploiting the receiver's multiple antenna capability of the SIMO system in overcoming the impact of channel impairments in AWGN channel models. The obtained results collected under SIMO are then compared to conventional SISO systems.
%
For each of the 10 simulated devices, whose hardware impairments values are shown in Table~\ref{Tb2}, the LLTF preamble of the IEEE 802.11n waveform is impaired with AWGN channels of SNR values ranging from -20 dB to 20 dB and then received and sampled at the receiver by its multiple receiving antennas. Once received, the waveform is averaged using the method described in Section~\ref{subsec:AWGN}, and the averaged signal is used to create the dataset used for training, testing, and validation as previously described.
\comment{
For this evaluation, the following steps were taken:
\begin{itemize}
  \item Different SIMO systems with numbers of receiving antennas varied from 2 to 10 and different SNRs with values varied from -10 dB to 20 dB are considered.
  \item Different CNN models were trained using data collected from SISO and SIMO systems over AWGN channels with SNR values ranging from from -20 dB to 20 dB separately.
   \item The models trained at 20 dB for SISO and SIMO systems were tested on data collected from channels with the same SNR value (same channel for training and testing), as well as channels with different SNR values (different channels for training and testing) to investigate the effect of channel noise and the number of receiving antennas on the classification accuracy.
\end{itemize}
}

 \begin{figure}
\centering
\includegraphics[width=.99\columnwidth]{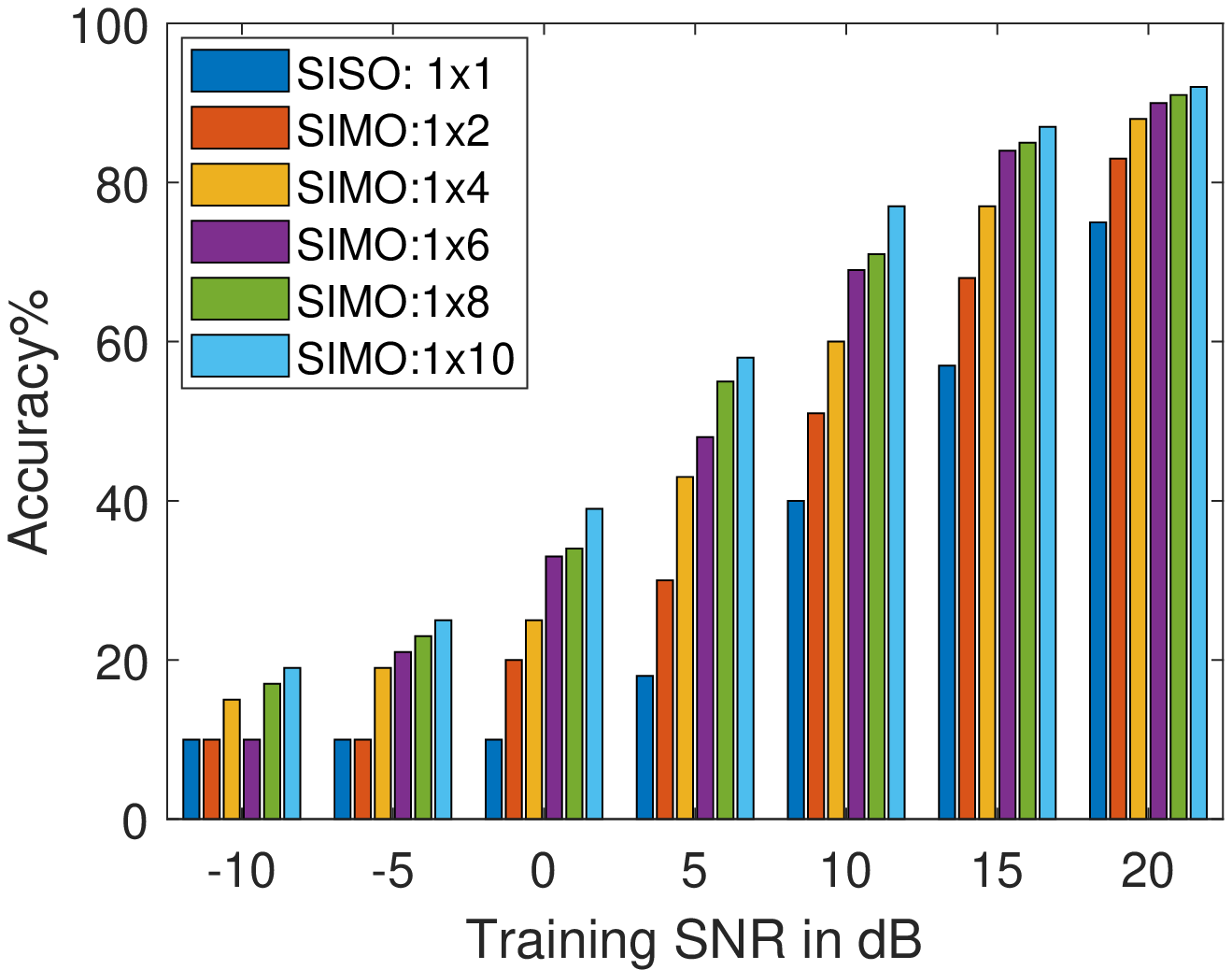}
\caption{Training accuracy when using the same AWGN channel for training and testing.}
\label{subfig:AWGN3}
\end{figure}
%

\begin{figure}
\centerline{
    \subfigure[Testing accuracy]
   {\includegraphics[width=0.5\columnwidth]{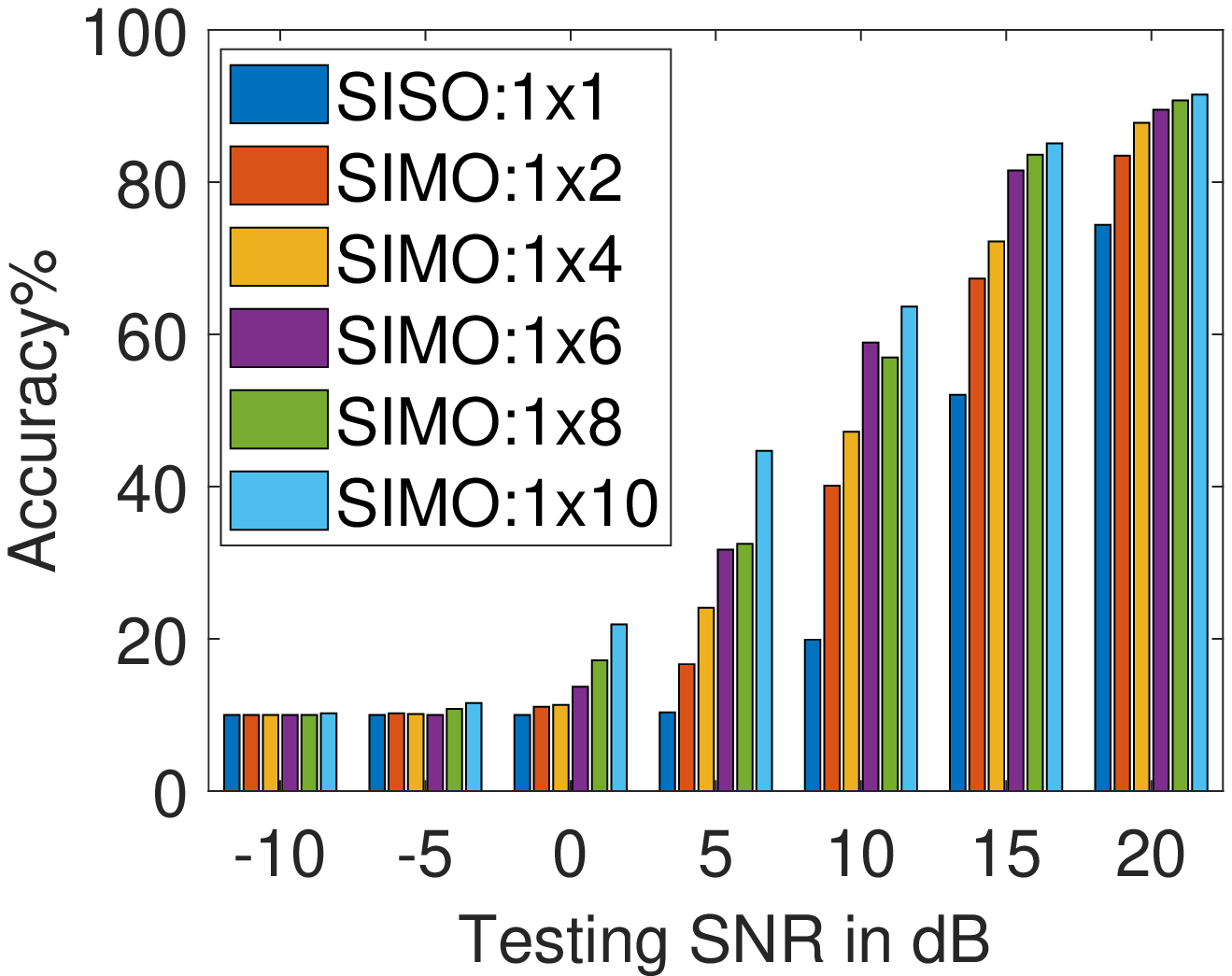}
    \label{subfig:AWGN1}}
    %
    \subfigure[SIMO over SISO accuracy gain]
    {\includegraphics[width=0.5\columnwidth]{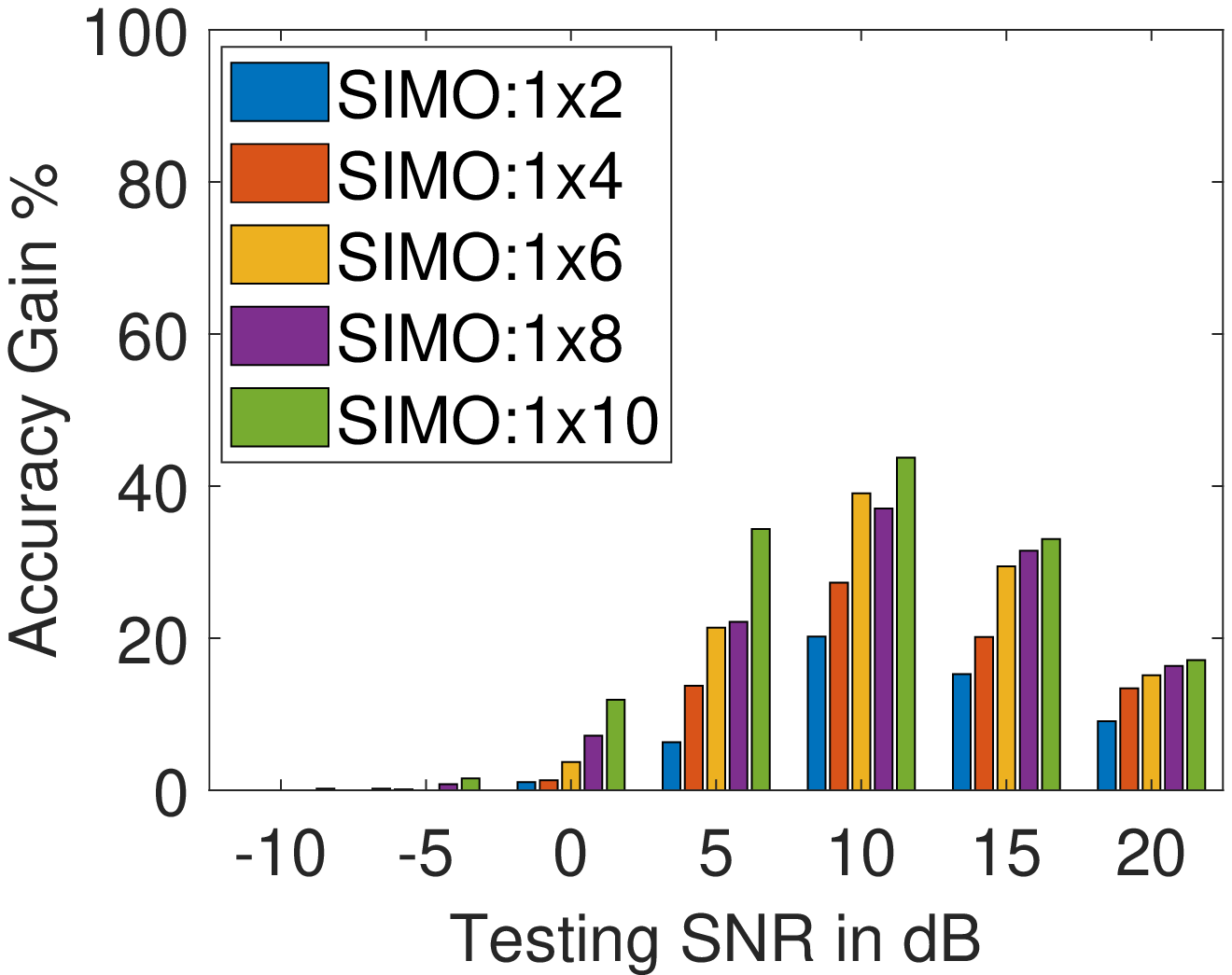}
    \label{subfig:AWGN2}}}
\caption{Testing accuracy when using different AWGN channels for training and testing. Training channel SNR is fixed at 20dB while testing SNR is varied from -10dB to 20dB.}
\label{fig:AWGN Results}
\end{figure}

\subsubsection{Same Training and Testing AWGN Channels}
Figure~\ref{subfig:AWGN3} shows the training accuracy obtained under different SNR values for SIMO links with varying number of receiving antennas. First and as expected, the figure shows that the worse the AWGN channel (i.e., the smaller the SNR value), the less accurate the CNN training outcome. For instance, a SIMO system with 2 receiving antennas achieves a training accuracy of $84\%$ at 20 dB, but only a $51\%$ accuracy at 10 dB.
Second, it is also observed that increasing the number of receiving antennas yields higher training accuracy when considering the same SNR. For example, training a 4-antenna SIMO system at 15 dB increases the training accuracy from $57\%$ to $77\%$ compared to the SISO system, whereas, a 10-antenna SIMO system increases it from $57\%$ to about $87\%$. Third, the figure also shows that the improved training accuracy of SIMO systems over the SISO system is more significant at SNR values ranging from 15 dB to 0 dB. However, at severe noisy channels with SNR values less than 0 dB, increasing the number of receiving antennas from 1 to 10 does not achieve a training accuracy higher than $40\%$ reflecting that the SNR gain acquired from combining the multiple received signals is no longer compensating the noisy channel effect and that the impairments are significantly overshadowed by the noise.

\subsubsection{Different Training and Testing AWGN Channels}
In Figure~\ref{subfig:AWGN1}, we show the testing accuracy obtained under different SNR values also while considering a SIMO system with a varying number of receiving antennas. Here, CNN models are trained at SNR = 20 dB, but tested at different SNRs, with SNRs being varied from -10 dB to 20 dB. We make three observations from this figure.
First, we observe that a decrease in the SNR value of the testing channel results in a decrease in the classification accuracy, and such a decrease is more profound for lower SNR values. This is true regardless of the number of receiving antennas. This observation clearly shows again the challenging impact that channel quality variations have on the classification accuracy, where achieving high accuracy when the channel used for testing has the same quality as that used for training does not guarantee that similar accuracy performance will be obtained when the channel used for testing is different from that used for training. However, such a discrepancy is less significant under the proposed SIMO approach, where the leveraging of multiple receiving antennas plays a key role in maintaining high accuracy even in the presence of channel quality degradation. For example, while testing over a 10 dB channel makes the SISO system accuracy drop from about $75\%$ to about $20\%$, it makes a SIMO system with 6 receiving antennas drop from about $90\%$ to about $59\%$ only. This demonstrates the benefit of SIMO in overcoming the impact of channel quality variations, and in making our proposed technique more channel agnostic.
Second, the figure also shows that the higher the number of receiving antennas is, the higher the achieved accuracy is, and this is regardless of the testing SNR. For example, at a tested channel of 15 dB, a SIMO system with 2, 4, 6, and 10 antennas achieve classification accuracy of about $67$, $72$, $82$ and $85\%$, respectively.
Third, we observe that for severely degraded channels (i.e., low SNRs), the CNN models perform poorly regardless of the number of received antennas, and the achieved classification accuracy is as good (or as bad) as what random classification would be. This can clearly be seen from Figure~\ref{subfig:AWGN1}, where the accuracy is similarly low for all studied systems.

Now in Figure~\ref{subfig:AWGN2}, we capture and depict the testing accuracy improvement/gain that the SIMO approach makes over the SISO/conventional approach. 
This figure demonstrates three key trends. 
First, we observe that the SIMO approach improves the robustness of the CNN models in terms of maintaining reasonably high accuracy even in the presence of channel variations (i.e., testing occurs on a channel different that is from that used for training). 
Such accuracy gains are especially significant for reasonably high SNRs, i.e., for SNR values above 5 dB. For instance, the SIMO approach improves the accuracy by up to $45\%$ even when testing occurs over a channel that is degraded by 10 dB.
Second, the figure shows that the improvement in accuracy made by the SIMO approach peaks at 10 dB, and not at the SNR value of the channel used during training; i.e., 20 dB.
Third, this gain in the testing accuracy obtained by increasing the number of receiving antennas vanishes when the SNR value used for testing drops below 5 dB. In other words, for severely degraded channels, neither the SISO approach nor the SIMO approach can maintain good enough classification accuracy.

The observed accuracy gain is achieved by the increased SNR value resulting from the averaging method used by the proposed SIMO approach as explained in Section~\ref{subsec:AWGN}. Such an SNR gain mitigates the noise effect allowing the CNN models to identify devices even in the presence impairments in the waveform. However, for severe noisy channels where the SNR gain obtained from increasing the number of receiving antennas is no longer enough to mitigate the increase in the noise power, the SIMO approach performance degrades and shows no improvement over the SISO system.
%
%
The results for both training and testing are commensurate, as both indicate that the averaging method using SIMO systems increases the SNR of the averaged signal as previously explained. Thus the CNN is able to capture the impairments effects on the waveform more efficiently despite the noise.

\subsection{Flat Fading Channel Results}
In order to ensure that the flat fading channels are blindly identifiable from the received signal up to a single complex scalar ambiguity as explained in Section~\ref{subsec:Fading}, we simulate and collect data from 10 devices each using a $3\times3$ MIMO system to send QPSK symbols encoded via Tarokh STBC $\mathcal{O}_{3}$~\cite{tarokh_space-time_1999-1}. 
This setting ensures that the channel is blindly identifiable even when the channel is rank deficient.
MATLAB 2020b communication toolbox is used to establish the aforementioned MIMO wireless link for each of the 10 devices. The transmitted signal blocks are impaired with a flat fading Rayleigh MIMO channel. At the receiving end, blind channel estimation is performed, and the estimated QPSK symbols are used for training and testing the CNN model.
\comment{
In this experiment, we evaluate the proposed MIMO-enabled approach and compare it to the conventional SISO approach, and we do so as follows.
\begin{itemize}
  \item We collect data from the MIMO wireless link for channels with different average path gain (APG) and maximum Doppler shift (MDS) values.
  \item The CNN was trained on data from a flat fading channel with training APG of -20 dB and training MDS of 0 Hz, and then tested on (i) dataset obtained over the same channel used for training (same channel for training and testing) and (ii) dataset obtained over channels with different testing APG values (different channels for training and testing).
  \item The CNN was trained on data from a flat fading channel with training APG of -20 dB and training MDS of 1/10 Hz, then tested on data from the same channel used for training (same channel for training and testing), and on data from channels with different testing MDS values (different channels for training and testing). 
   \item For completion, the previous two steps were repeated for different APG and MDS values ranging from -20 to 20 dB and from 1/2000 to 1 Hz, respectively.
 \end{itemize}
 }

\subsubsection{Same Training and Testing Rayleigh Channels}
\begin{figure}
\centerline{
    \subfigure[Training/testing MDS= 0Hz]
   {\includegraphics[width=.5\columnwidth]{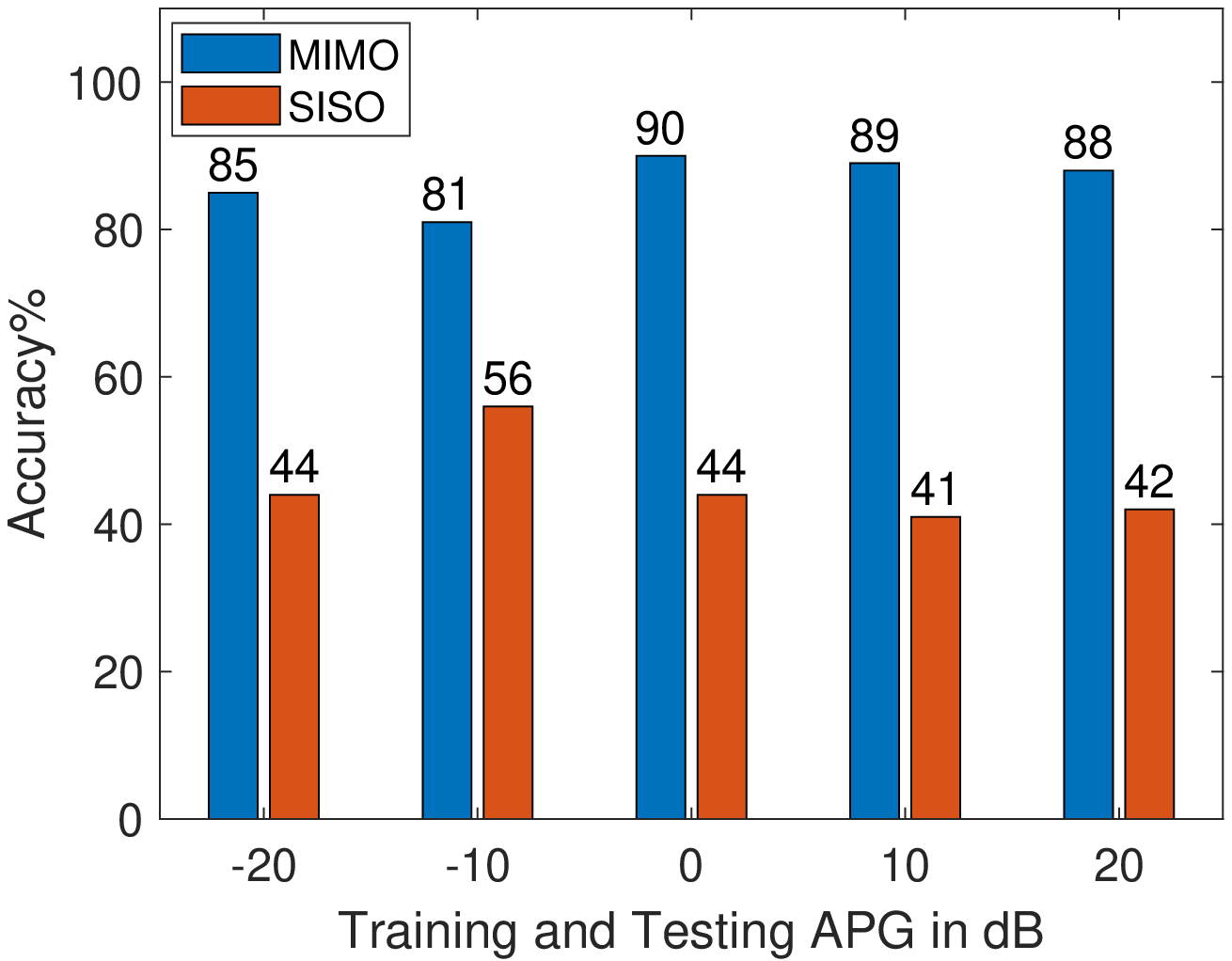}
    \label{subfig:same-APG}}
    \subfigure[Training/testing APG= -20dB]
    {\includegraphics[width=.5\columnwidth]{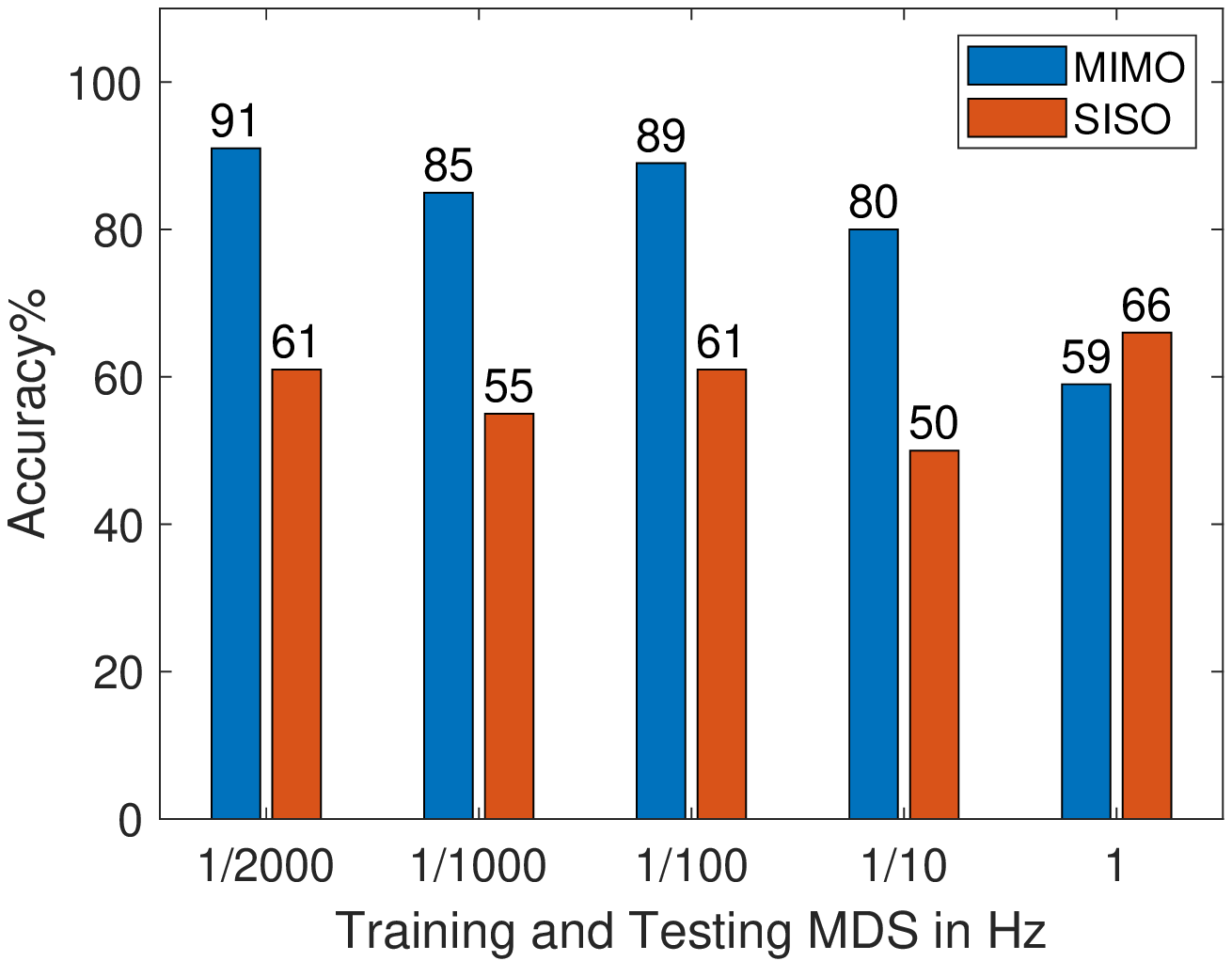}
    \label{subfig:same-MDS}}}
\caption{Testing accuracy when the same Rayleigh channel is used for training and testing.} 
\label{fig:same channel}
\end{figure}
In this section, we analyze the results obtained when training and testing are performed over the same Rayleigh fading channel for both MIMO and SISO approaches.
Figure~\ref{fig:same channel} shows the obtained testing accuracy while varying the APG (Figure~\ref{subfig:same-APG}) and MDS (Figure~\ref{subfig:same-MDS}) values of the channel.
This figure clearly shows that the proposed MIMO-enabled approach increases the testing accuracy significantly compared to the conventional/SISO approach. %
For instance, from Figure~\ref{subfig:same-APG}, which captures the impact of APG on the obtained accuracy, we observe that at training and testing APG of -20 dB, the MIMO-enabled approach increases the testing accuracy from $44\%$ to up to $85\%$ compared to the SISO approach, thereby doubling the obtained accuracy.
In addition, we observe that the accuracy improvement that MIMO-enabled approach offers over the conventional approach is consistent across the entire APG value range, i.e., our proposed MIMO-enabled classification approach doubles the accuracy regardless of the APG value.

Now when we look at Figure~\ref{subfig:same-MDS}, which depicts the impact of MDS on the obtained accuracy, still when the same channel is used for both training and testing,
we first observe that the MIMO-enabled approach outperforms the SISO approach at MDS values less than 1 Hz. For example, at MDS of 1/10 Hz, MIMO achieves a testing accuracy of $80\%$ whereas the SISO/conventional approach achieves only $50\%$. However, at MDS of 1 Hz, both approaches result in a degraded testing accuracy of about $60\%$. This result indicates that the proposed MIMO-based approach is effective over the SISO approach for slow channels.
Second, we observe that an increase in the MDS value results in a decrease in the testing accuracy for the MIMO-based approach. For instance, at MDS = 1/10 Hz, the achieved accuracy is $80\%$ and at MDS = 1 Hz, the accuracy drops to $59\%$. However, for the SISO approach, the testing accuracy fluctuates from $50\%$ to $66\%$.
\begin{figure*}
\centerline{
    \subfigure[Training APG = 20dB]
   {\includegraphics[width=0.4\columnwidth]{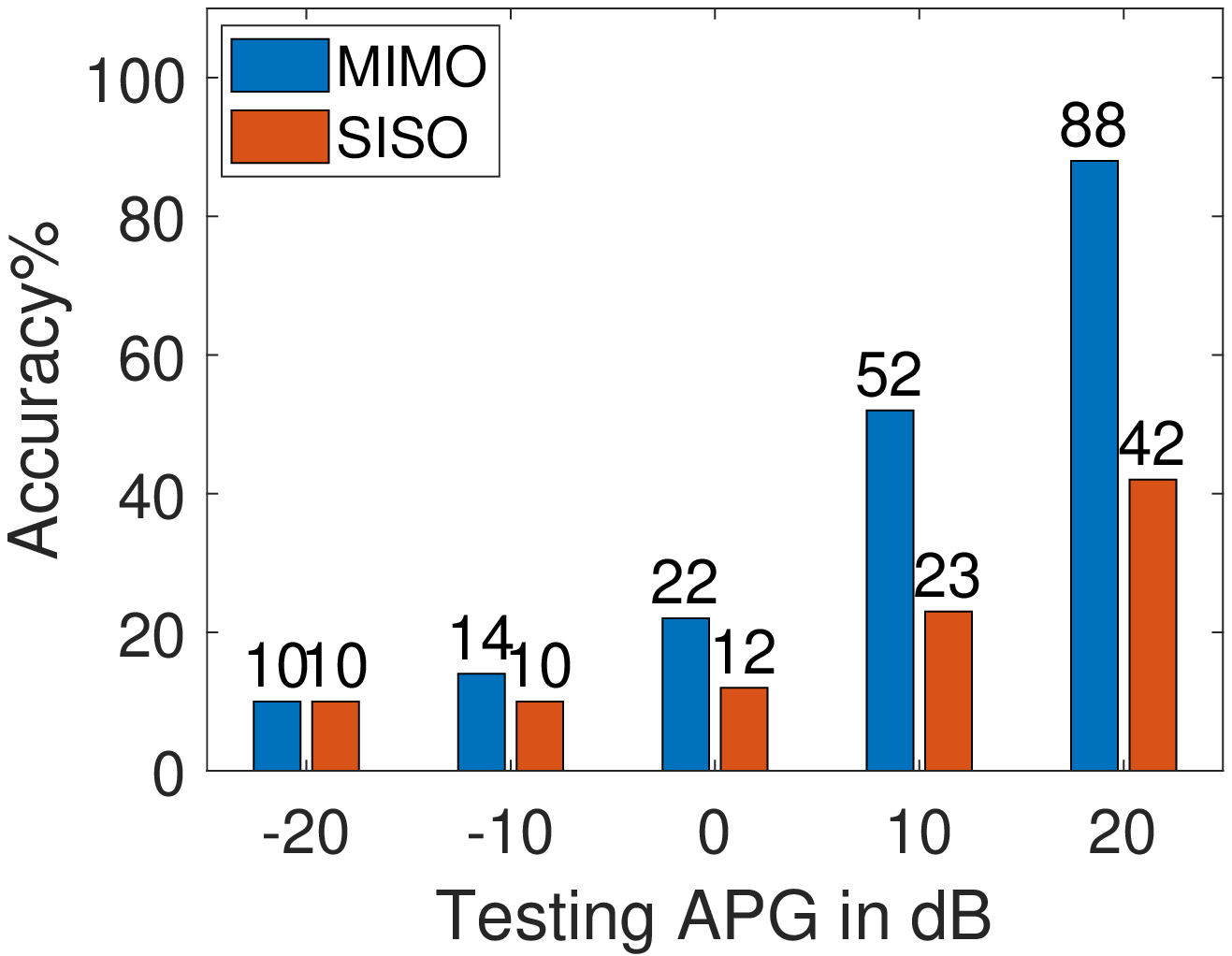}
    \label{subfig:fading1}}
    \subfigure[Training APG = 10dB]
   {\includegraphics[width=0.4\columnwidth]{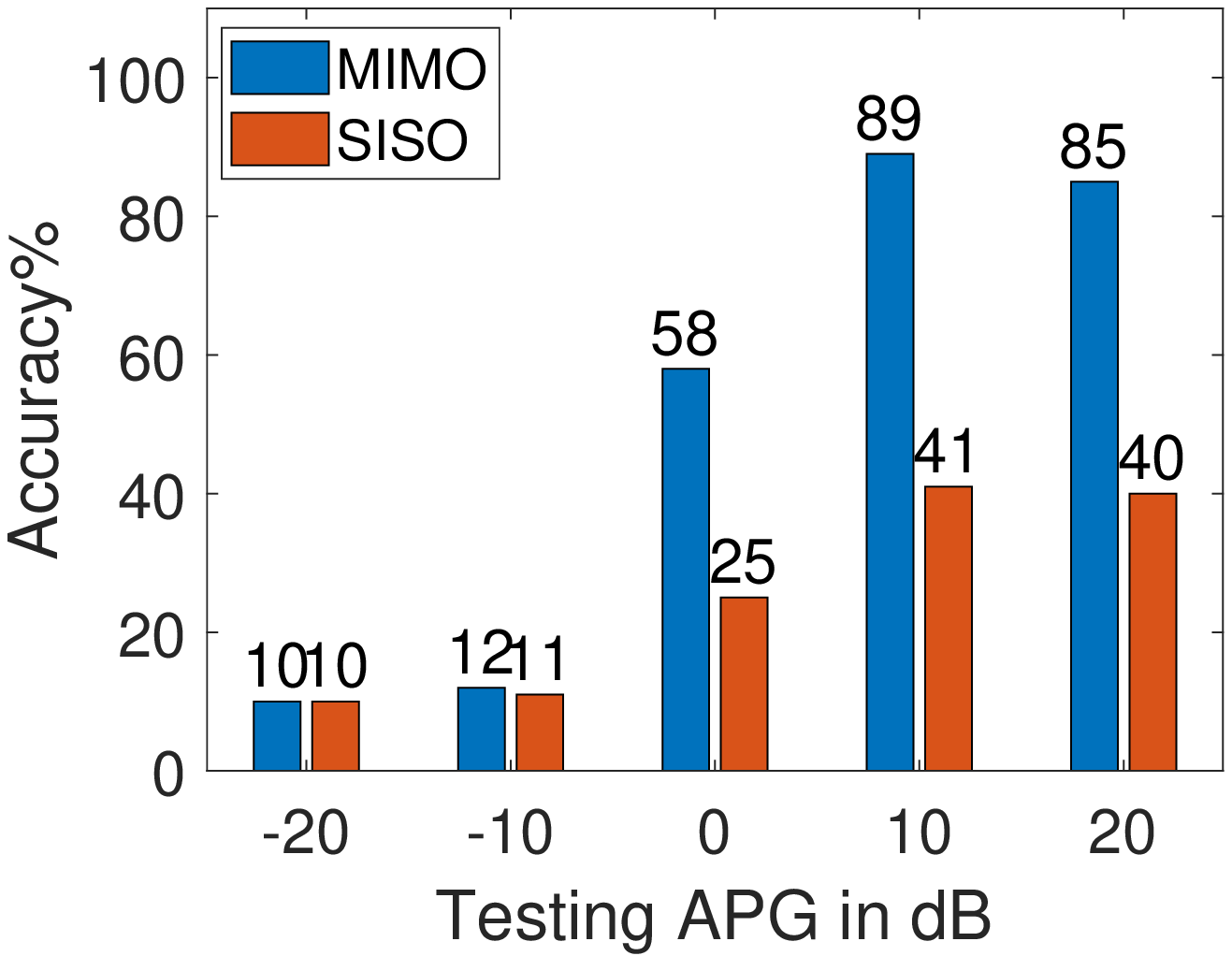}
    \label{subfig:fading2}}
    %
    \subfigure[Training APG = 0dB]
    {\includegraphics[width=0.4\columnwidth]{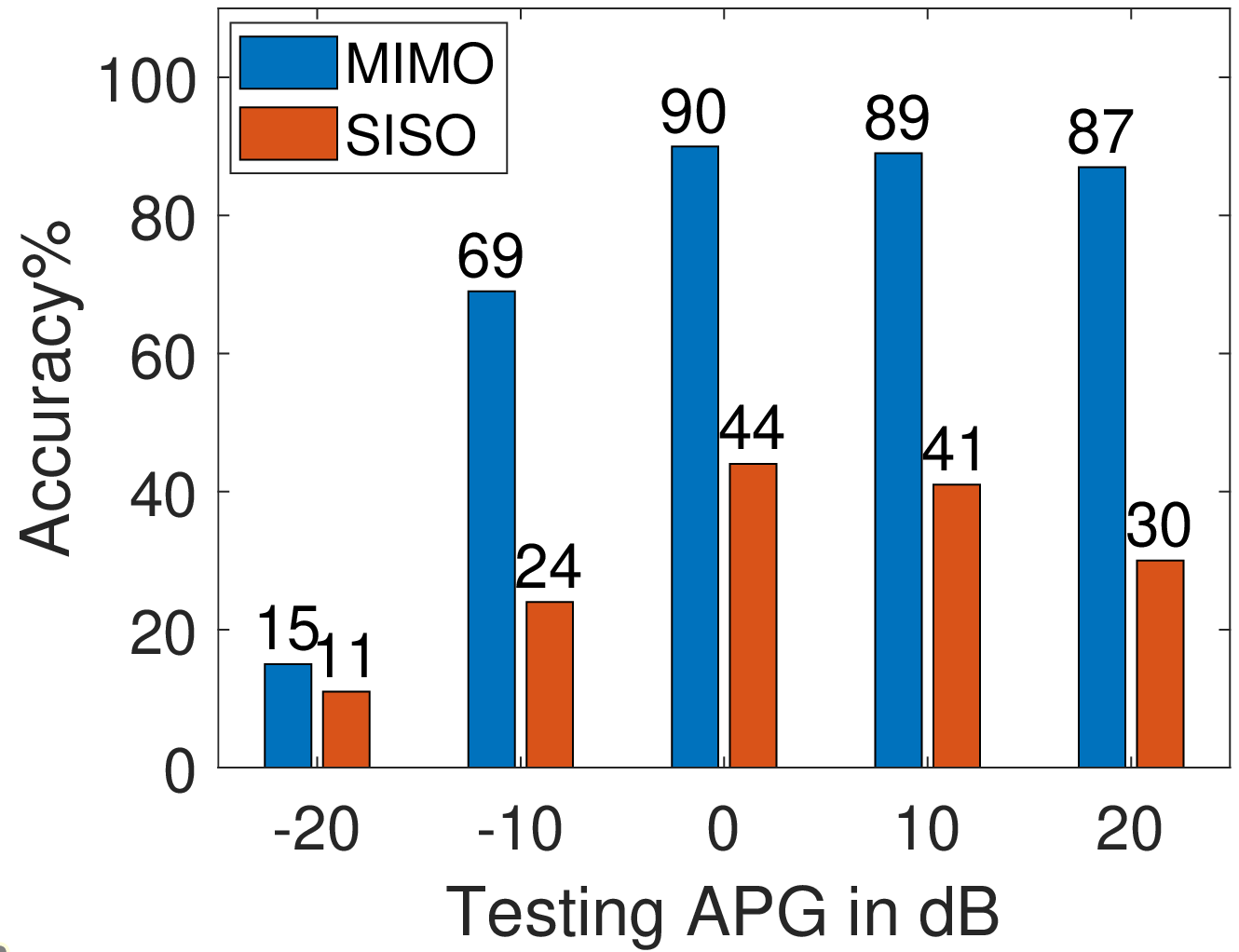}
    \label{subfig:fading3}}
    %
    \subfigure[Training APG = -10dB]
    {\includegraphics[width=0.4\columnwidth]{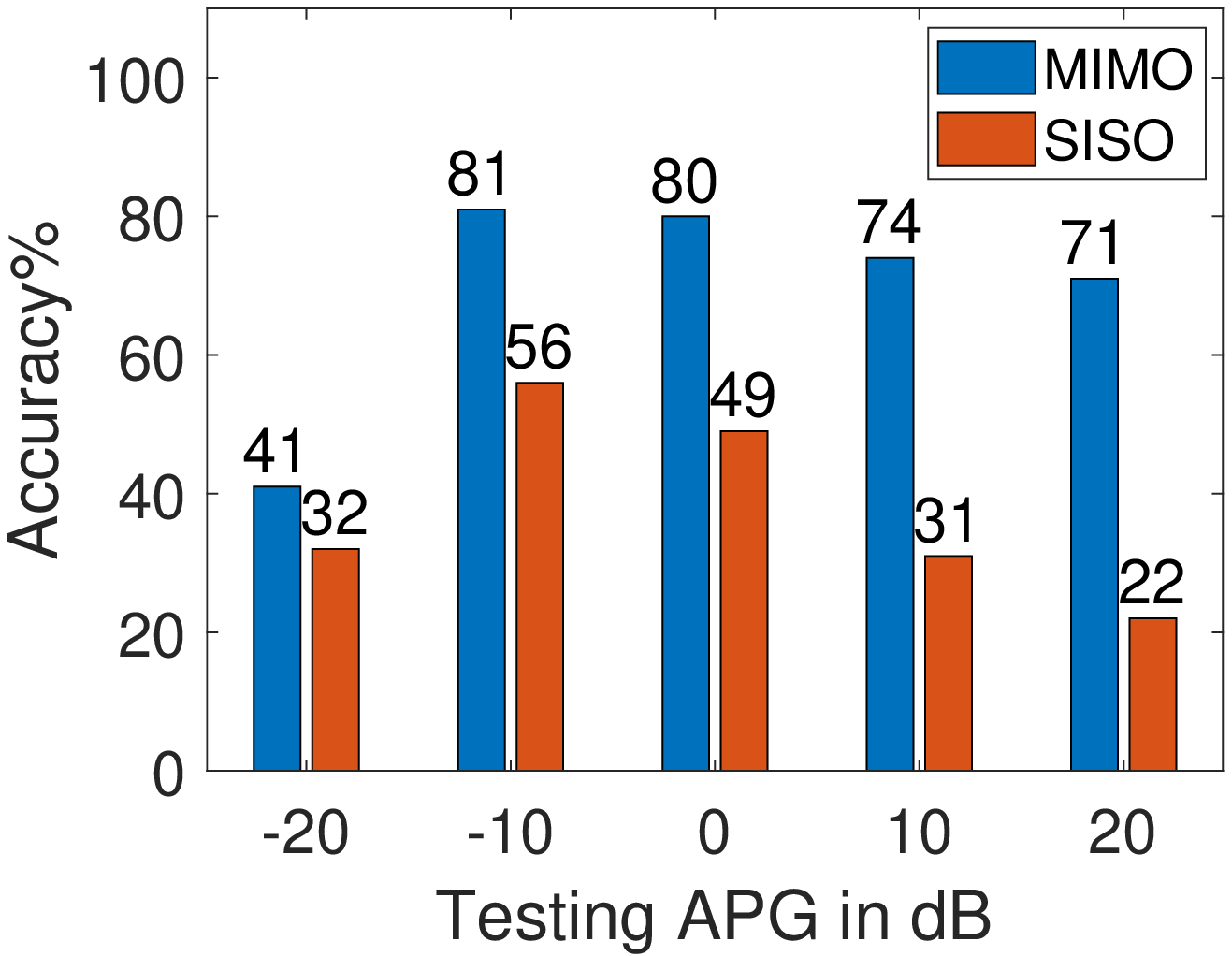}
    \label{subfig:fading4}}
    %
    \subfigure[Training APG = -20dB]
    {\includegraphics[width=0.4\columnwidth]{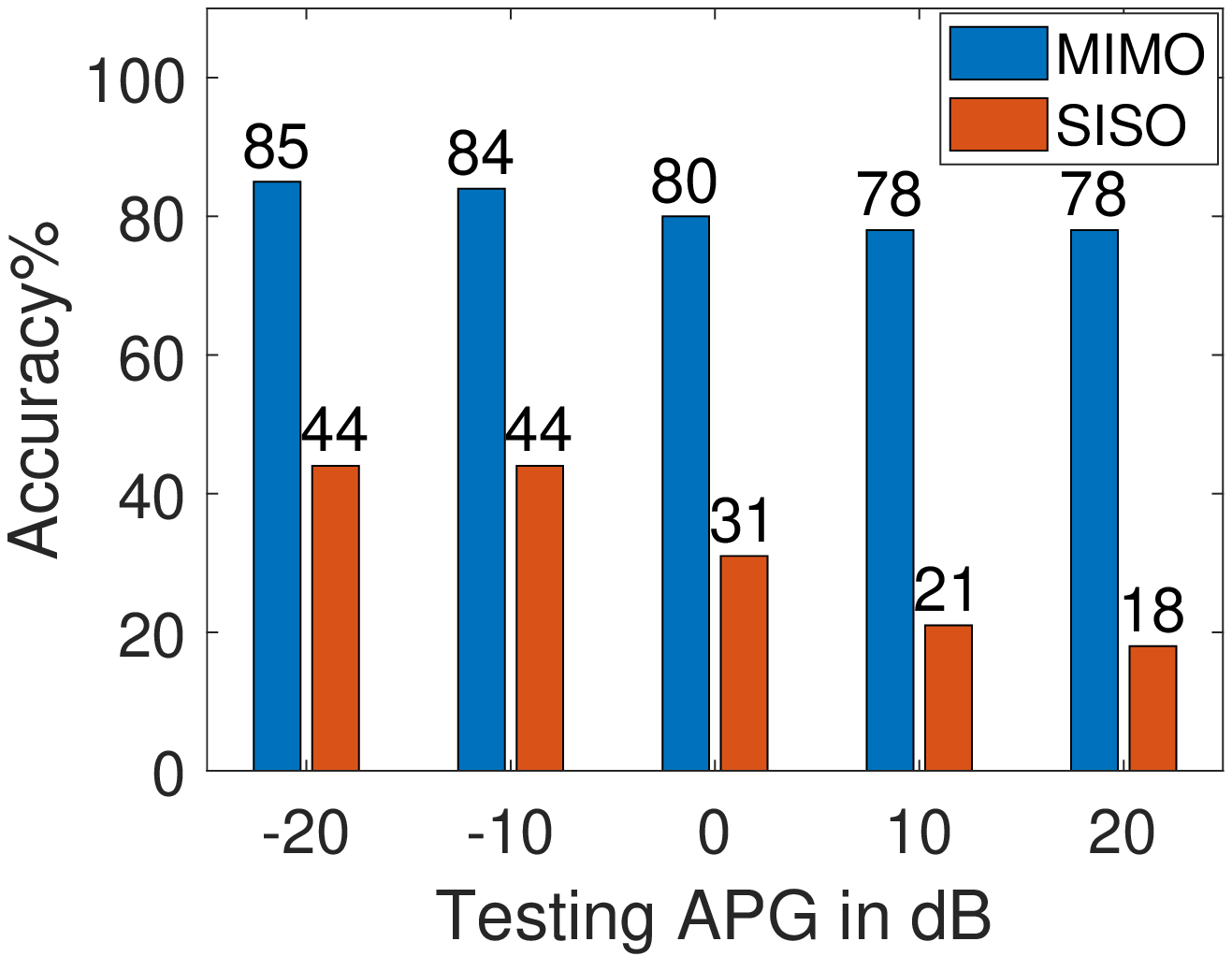}
    \label{subfig:fading5}}}
\caption{Impact of APG on accuracy. MDS is fixed at 0 Hz.} 
\label{fig:diff-apg}
\end{figure*}

\subsubsection{Different Training and Testing Rayleigh Channels}
We now present, analyze and compare the results obtained when the channel used for training is different from that used for testing. This scenario mimics real world settings where the classification models are trained under certain channel conditions, but then used later for real-time device classification under possibly different channel conditions.
\paragraph{Impact of APG: Testing Accuracy}
%
\begin{figure}
\centering
\includegraphics[width=1\columnwidth]{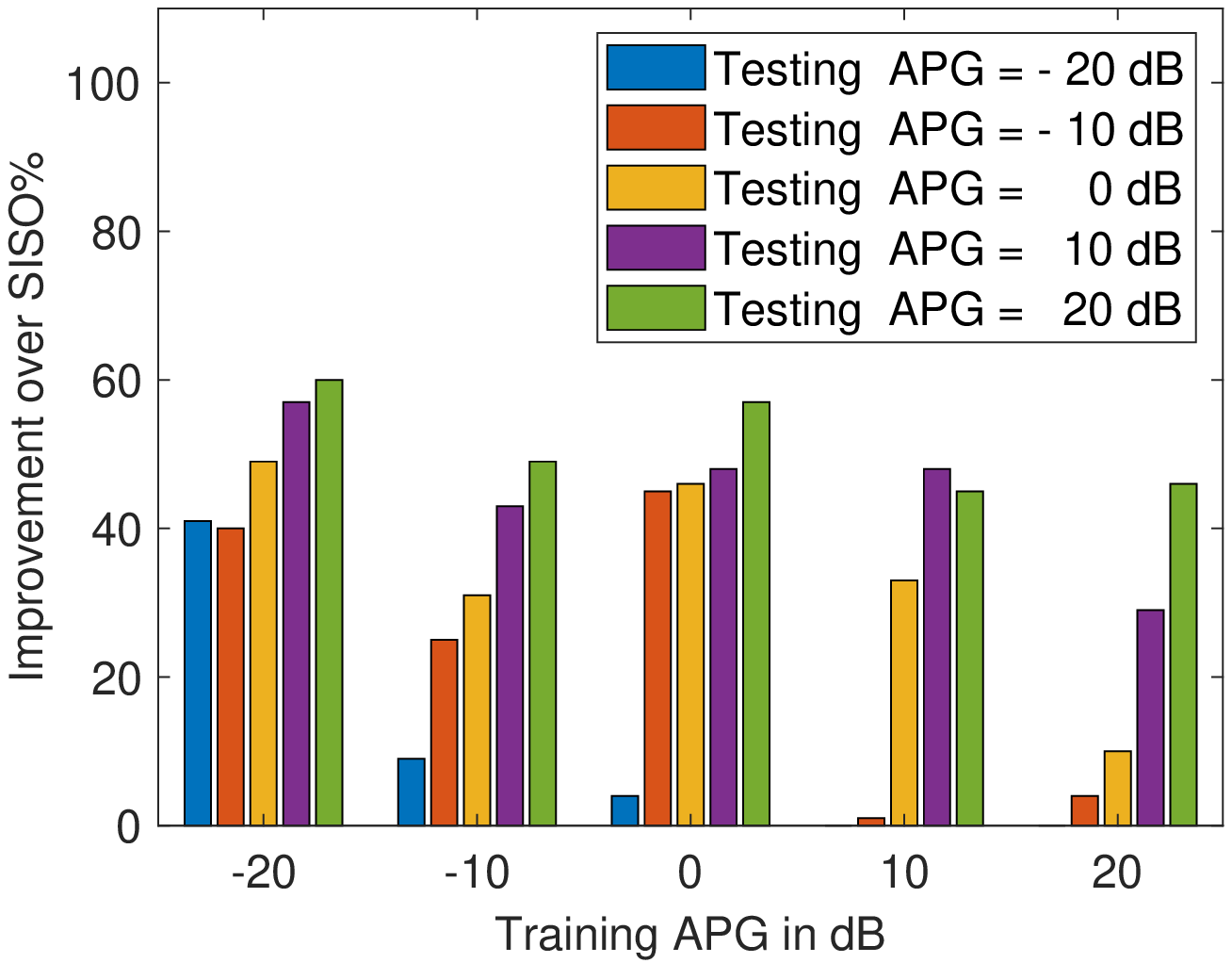}
\caption{Impact of Training APG on the MIMO improvement over SISO under flat fading Rayleigh channels}
\label{Improvement over SISO}
\end{figure}
Figure~\ref{fig:diff-apg} shows the testing accuracy over Rayleigh channels where the models are trained and tested over channels with different APG values, ranging from 20 to -20 dB. For this figure, MDS is set to 0Hz.
This figure shows the effect of the wireless channel on the device classification accuracy. For instance, we observe that when the conventional/SISO method is used and the models are trained over channel with training APG of -20 dB and tested over a channel with testing APG of 20 dB (Figure~\ref{subfig:fading5}), the accuracy drops from $44\%$ to $18\%$. Our observation indicating the seriousness of the channel impact on device classification accuracy is well aligned with previous work findings as discussed in Sections~\ref{sec:Introduction} and~\ref{sec:related Works}.
The figure also shows that the proposed MIMO-enabled approach overcomes significantly the impact of channel variations by improving the testing accuracy over the conventional/SISO approach when the CNN is trained and tested on different channels, and this especially true when the training channels exhibit severe fading. In addition, we observe that the MIMO-based approach testing accuracy is more stable when the CNN is trained at severe fading channels. 
Looking at Figure~\ref{subfig:fading1}, which depicts the testing accuracy under varied testing APG values of the testing channel while fixing the training APG to 20 dB,
we observe that for testing APG greater that 0 dB, MIMO achieves improved performance over SISO. However, when the testing APG is less than 0 dB, the testing accuracy is unreliable, and both SISO and MIMO approaches are equivalent. For instance, when training APG is 20 dB and testing APG is 10 dB, the MIMO-based approach achieves a testing accuracy of about $52\%$, but only about $23\%$ is achieved under the conventional/SISO approach. However, when the training APG is 20 dB and the testing APG is -10 dB, both MIMO and SISO approaches provide severely degraded and unreliable testing accuracy.
Now in Figure~\ref{subfig:fading2}, which depicts the testing accuracy under varied APG values of the testing channel while fixing APG of training channel to 10 dB, we observe that
the MIMO-based approach achieves significant higher testing accuracy compared to the SISO approach when the testing APG is greater than 0 dB. For instance, when the training APG is 10 dB and the testing APG is 20 dB (channel with less severe fading), MIMO system achieves a testing accuracy of about $85\%$, whereas only about $40\%$ is achieved under SISO. Moreover, when the testing APG = 0 dB (channel with more severe fading), the testing accuracy improves from $25\%$ to about $58\%$ when considering the MIMO-based approach versus the SISO approach.
Second, we observe that no matter how we change the testing channel, i.e. changing the channel for less/more severe fading channel, the testing accuracy degrades for both approaches. However, the original high training accuracy achieved by the MIMO-based approach enables it to achieve a higher testing accuracy for a wider range of testing APG when compared to SISO.

Figure~\ref{subfig:fading3} and Figure~\ref{subfig:fading4} capture the testing accuracy under varied APG values of the tested channel while fixing the training APG to 0 and 10 dB, respectively. From the two figures we observe the same trends of Figure~\ref{subfig:fading2}, and the MIMO-based approach achieves higher classification accuracy compared to the SISO/conventional approach when the testing APG is varied from the training APG.
Figure~\ref{subfig:fading5} captures the testing accuracy when varying the testing APG values at a fixed training APG of -20 dB.
From this figure we first observe that when testing on a different channels with testing APG values varying from -10 dB to 20 dB, the MIMO-based classification approach achieves an improved and stable testing accuracy when compared to the SISO approach. For instance, when the training APG is -20 dB and the testing APG is 20 dB, the MIMO approach achieves up to $60\%$ increase in the testing accuracy over the SISO approach.
Second, when the training APG is -20 dB (severe fading channel), the testing accuracy of the MIMO approach at different channels with testing APG varying from -10 dB to 20 dB does not go below $78\%$ compared to the SISO approach where the testing accuracy degrades to $18\%$. 

Figure~\ref{Improvement over SISO} illustrates the improvement/gain in testing accuracy that the MIMO-based approach achieves over the SISO approach on Rayleigh channels when the CNN is trained and tested under various different APG values, ranging from 20 dB to -20 dB. From this figure we can make the following observations.
First, when the CNN is trained on severe fading channel with training APG of -20 dB, the MIMO-based approach shows significant improvement over the conventional/SISO approach, and when the CNN is trained on less sever fading channels, i.e. training APG values higher than -20 dB , the improvement over the SISO approach decreases. For instance, an improvement of up to $60\%$ in the MIMO approach performance is achieved when the training APG is -20 dB and the testing APG values varies from -10 dB to 20 dB. However, when the training APG is -10 dB, the MIMO improvement decreases to $50\%$. This could be justified by the severely degraded testing accuracy for the SISO system over severe fading channels.
Second, we observe that when the CNN is trained at less severe fading channels with training APG values varying from 0 dB to 20 dB and tested at severe fading channels with testing APG values less than -10 dB, the MIMO system improvement over SISO vanishes. For example, when the training APG is 20 dB and the testing APG is -20 dB, both approaches show the same severe degraded testing accuracy.
Third, we observe that when the CNN is trained on severe fading channels then tested on less severe fading channels, the MIMO-based approach shows significantly improved performance over the SISO approach. Note that when the training APG is -10 dB and the testing APG is 10 dB, MIMO achieves about $43\%$ improvement over SISO, compared to only $9\%$ improvement when the testing APG is -20 dB. 
%
%
\begin{figure*}
\centerline{
    \subfigure[Training APG = 20 dB]
   {\includegraphics[width=0.4\columnwidth]{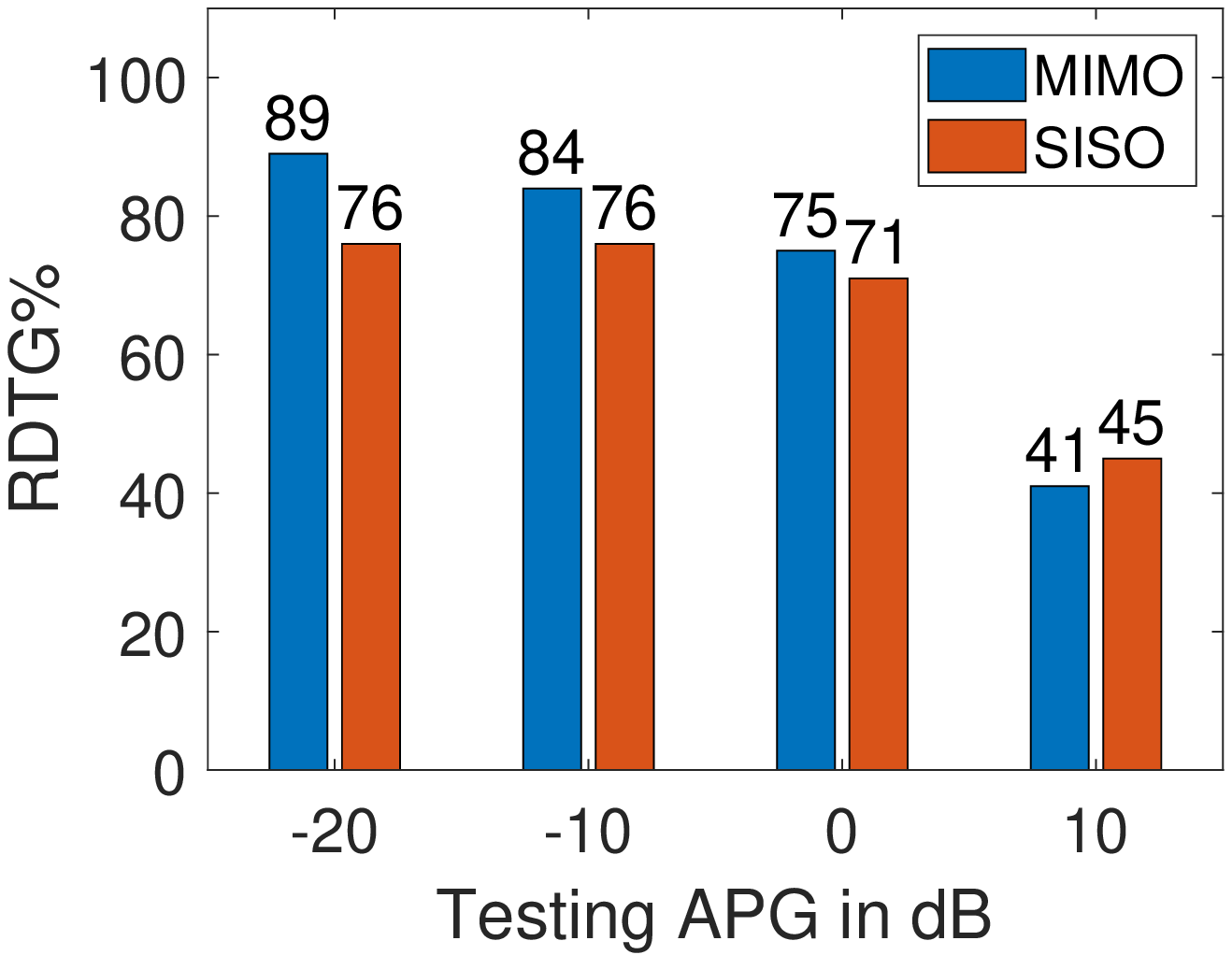}
    \label{subfig:RDTG1}}
    \subfigure[Training APG = 10 dB]
   {\includegraphics[width=0.4\columnwidth]{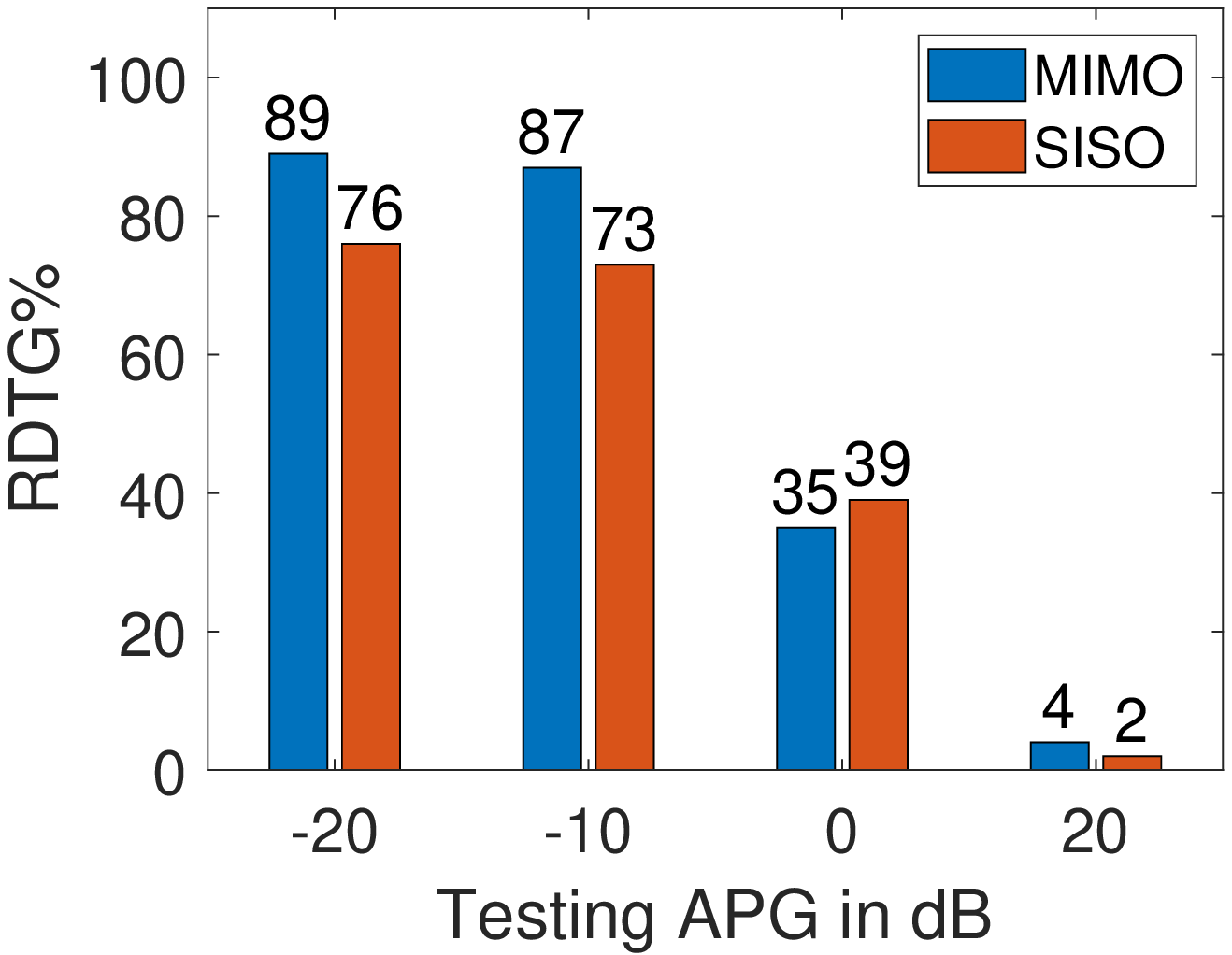}
    \label{subfig:RDTG2}}
    %
    \subfigure[Training APG = 0 dB ]
    {\includegraphics[width=0.4\columnwidth]{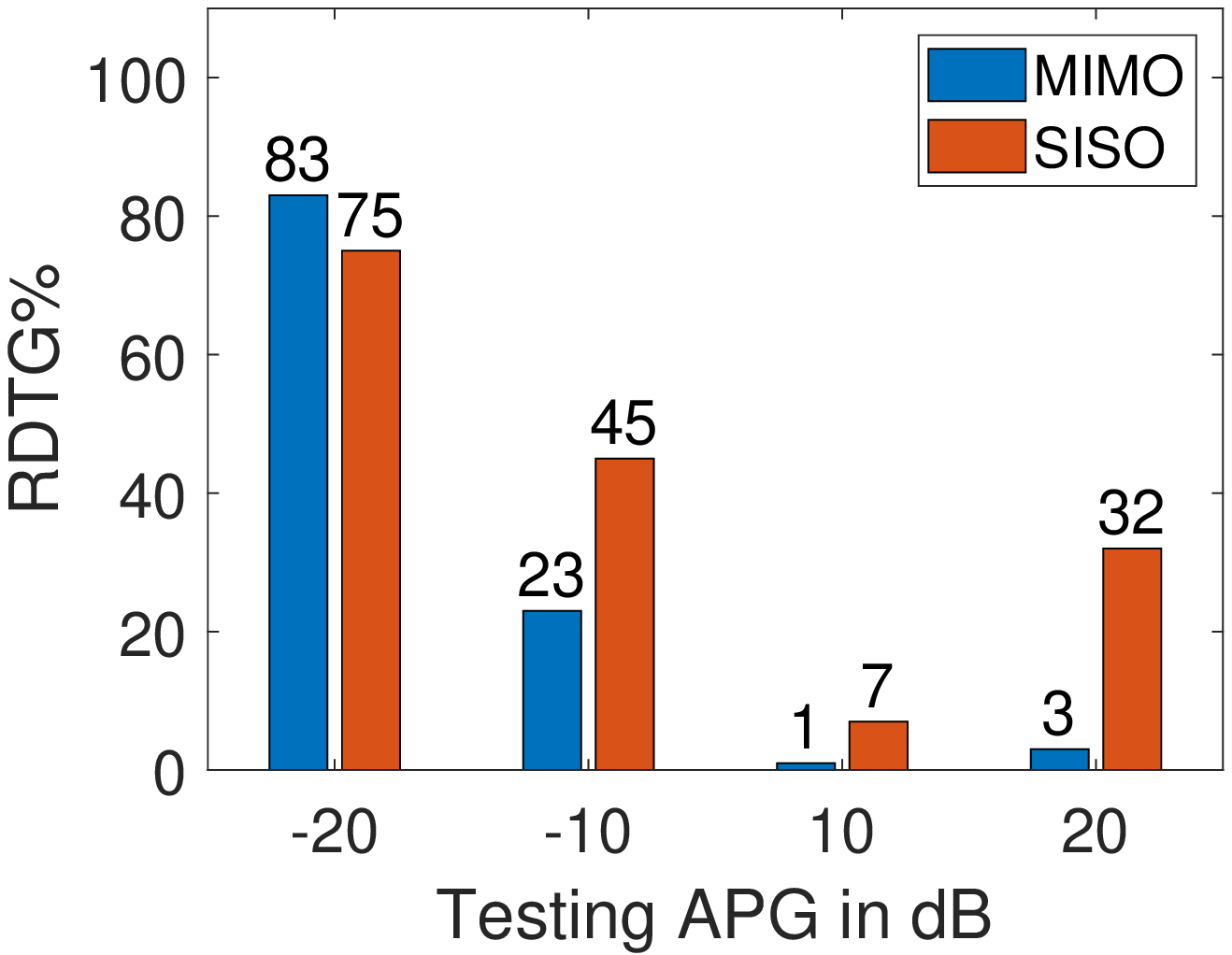}
    \label{subfig:RDTG3}}
    %
    \subfigure[Training APG = -10 dB]
    {\includegraphics[width=0.4\columnwidth]{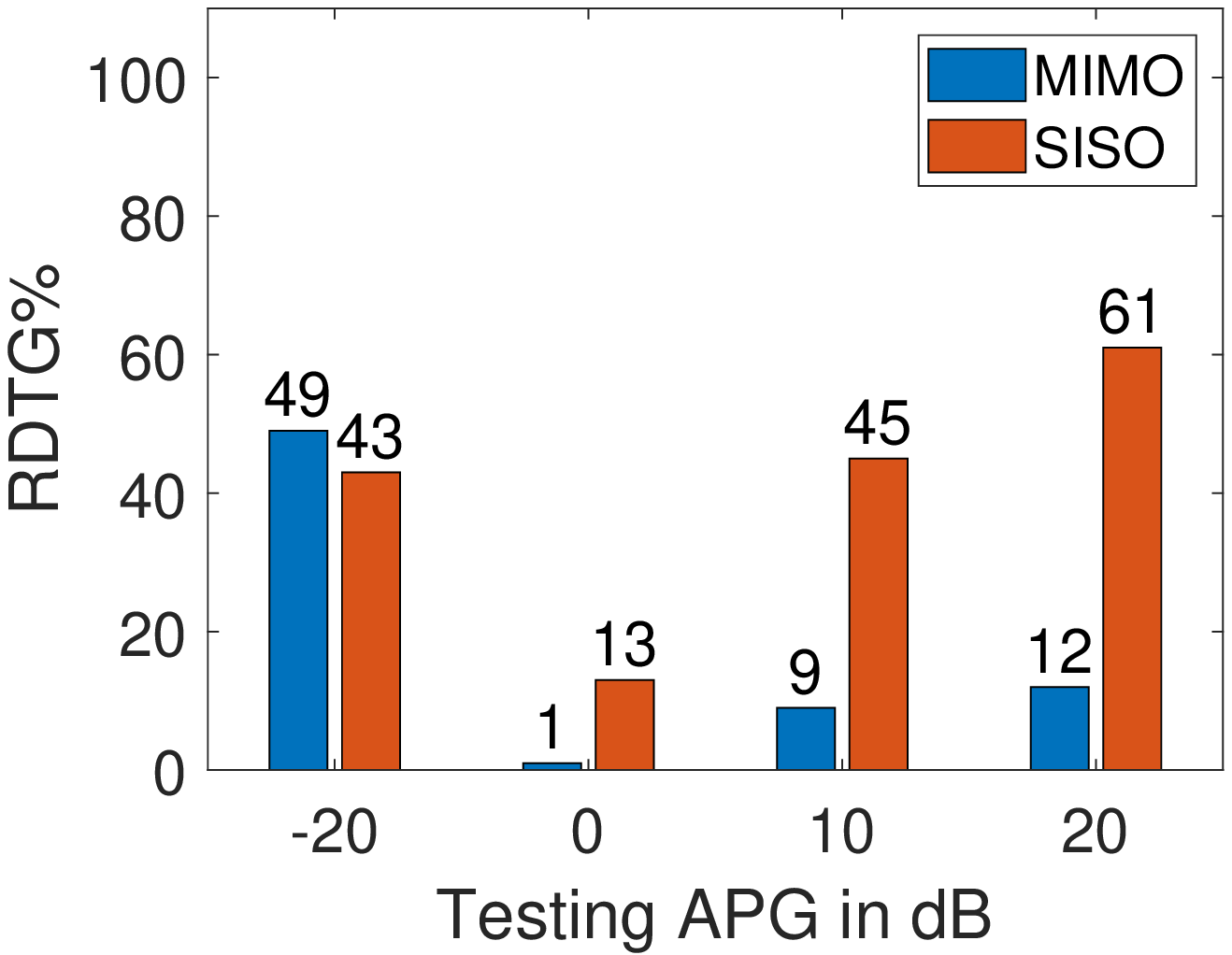}
    \label{subfig:RDTG4}}
    %
    \subfigure[Training APG = -20 dB ]
    {\includegraphics[width=0.4\columnwidth]{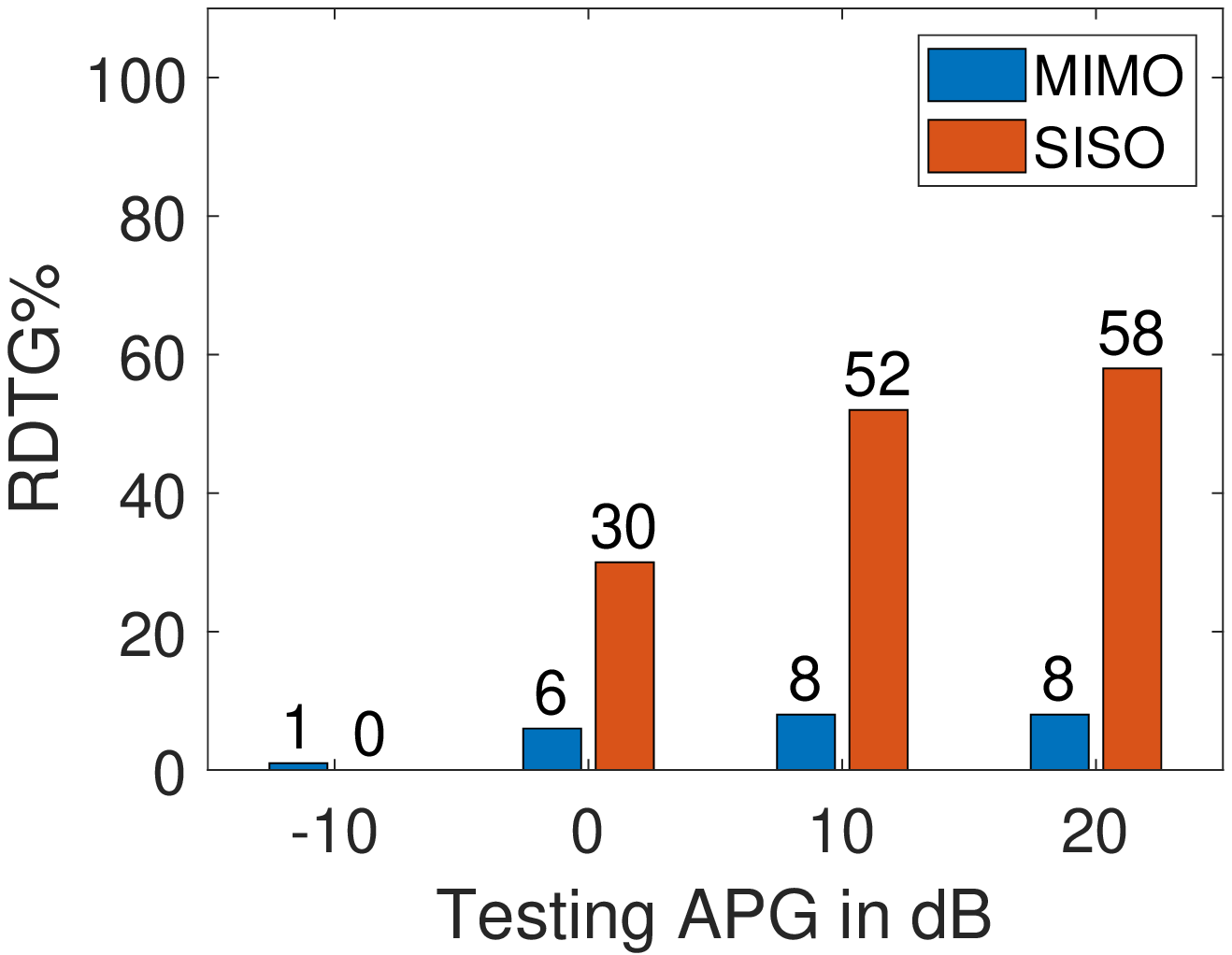}
    \label{subfig:RDTG5}}}
\caption{Impact of APG on RDTG. MDS is fixed at 0 Hz.} 
\label{RDTG}
\end{figure*}
\paragraph{Impact of APG: RDTG}
We now assess the robustness of the proposed MIMO-enabled fingerprinting approach against Rayleigh channel condition variations through the study of the RDTG performance metric. Recall that RDTG, as introduced in Section~\ref{subsec:metrics}, is defined as the percentage of reduction in the testing accuracy that occurred due to a change in the channel used for testing when compared to that used for training.
%
Figure~\ref{RDTG} shows RDTG values when training and testing of the learning models are done over Rayleigh channels with varying APGs but a fixed MDS. 
First, observe that compared to the SISO approach, the MIMO-based approach provides a much higher resiliency to channel variations, i.e. yields smaller RDTG values, when the training channel exhibits moderate (training APG = 0 dB; Figure~\ref{subfig:RDTG3}) to severe (training APG = -20 dB; Figure~\ref{subfig:RDTG5}) fading conditions. Observe that RDTG values can go as high as 61\% under the SISO approach, but does not exceed 12\% under the proposed MIMO-enabled approach.
%
One explanation for this observation is that despite the MIMO-enabled blind estimation, there remains some ambiguity in the estimated channel that affects the learning models in the training phase, where the models still learn features that are extracted from both channel and device impairments. Therefore, training at less severe flat fading channels while testing at more severe fading channels yields significant reduction in the accuracy. However, the MIMO-based approach still outperforms the SISO approach in spite of this unsolved ambiguity. 

Second, for high training APG values (e.g., 20 dB in Figure~\ref{subfig:RDTG1} and 10 dB in Figure~\ref{subfig:RDTG2}),
we observe that although the MIMO approach still outperforms the SISO approach in terms of the testing accuracy, the RDTG gap for SISO is less compared to the MIMO-based approach. For instance, in Figure~\ref{subfig:RDTG1}, when testing APG = -20 dB, the MIMO-based approach achieves an RDTG value of $89\%$ and the SISO approach achieves a value of $76\%$. This observation could be also explained by the effect of the remained ambiguity on the CNN as explained before.
Observe that the closer the testing APG values are to the training ones, the smaller the RDTG values achieved under MIMO are. This observation is commensurate with the previous observations made in Figure~\ref{fig:diff-apg} about the MIMO-based approach testing accuracy degradation when tested on more severe fading channels.
%

Now when considering moderate training APGs like 0 dB as in Figure~\ref{subfig:RDTG3}, we observe
that the MIMO-based approach is less immune to the channel variation when tested on a channel that is significantly more severe than that used for training. For instance, when testing APG = -20 dB, the RDTG value achieved under the MIMO-based approach is $83\%$. However, as shown in Figure~\ref{subfig:RDTG3}, the higher the testing APG, the lesser the RDTG value; i.e., the more resilient the MIMO-enabled approach is to channel condition variations. For example, when testing APG = -10 dB, the SISO approach achieves a RDTG value of $45\%$ compared to only $23\%$ achieved under the MIMO approach.
%

%
Figure~\ref{subfig:RDTG5}, depicting RDTG values when training APG = -20 dB,  
shows that when the fading conditions of the training channel become worse, the SISO approach continues to degrade significantly, but not so for the proposed MIMO approach. 
As an example, when testing APG = 0 dB, the SISO approach yields an RDTG value of $30\%$ whereas the proposed MIMO approach achieves an RDTG value of only $6\%$.
%
Moreover, as the testing APG continues to increase, while the MIMO-based approach maintains stable RDTG values (about $8\%$), SISO testing accuracy continues to degrade significantly, reaching RDTG values of up to $58\%$.

\begin{figure*}
\centerline{
    \subfigure[Training MDS = 1/2000 Hz]
   {\includegraphics[width=0.4\columnwidth]{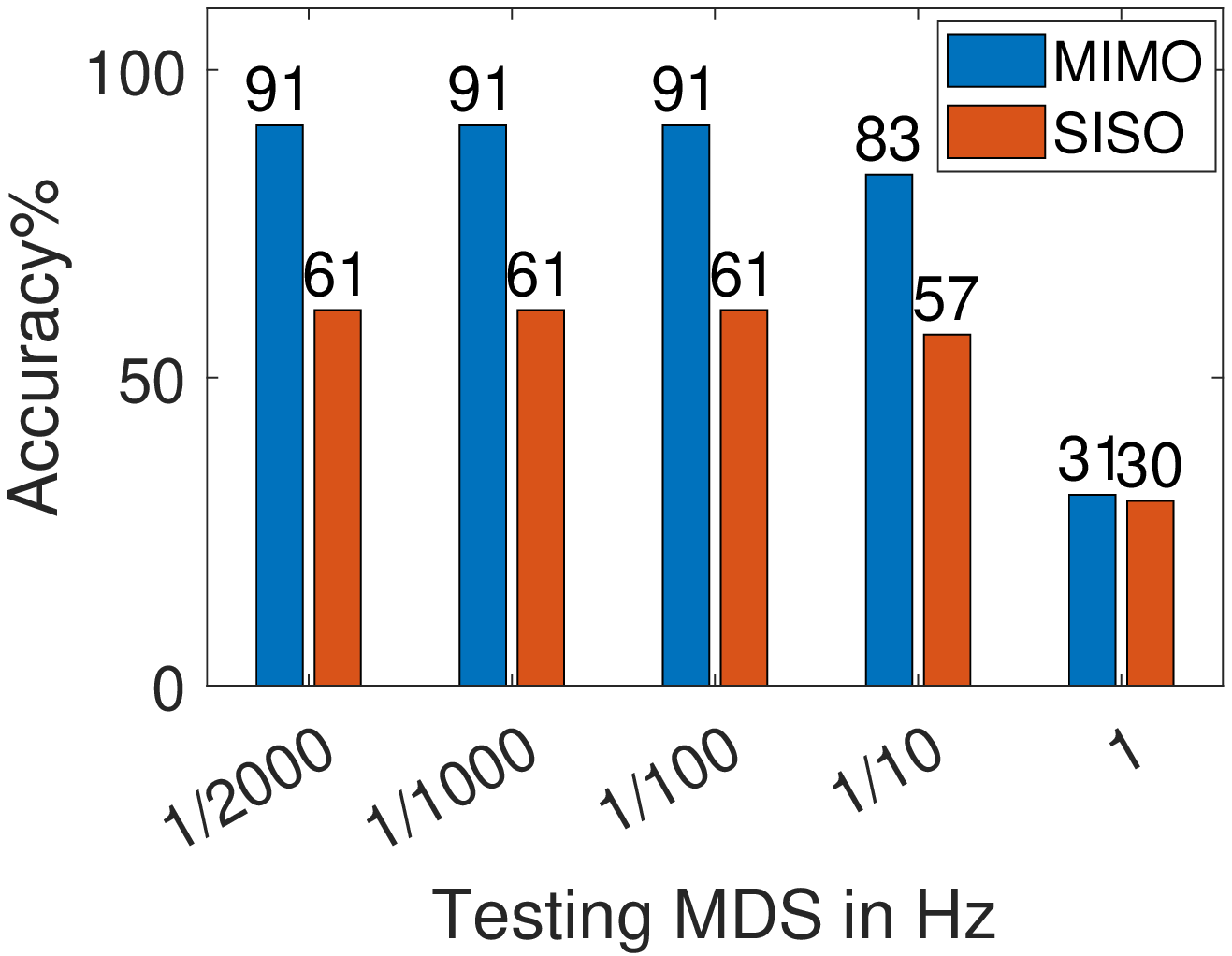}
    \label{subfig:fading6}}
    \subfigure[Training MDS = 1/1000 Hz]
   {\includegraphics[width=0.4\columnwidth]{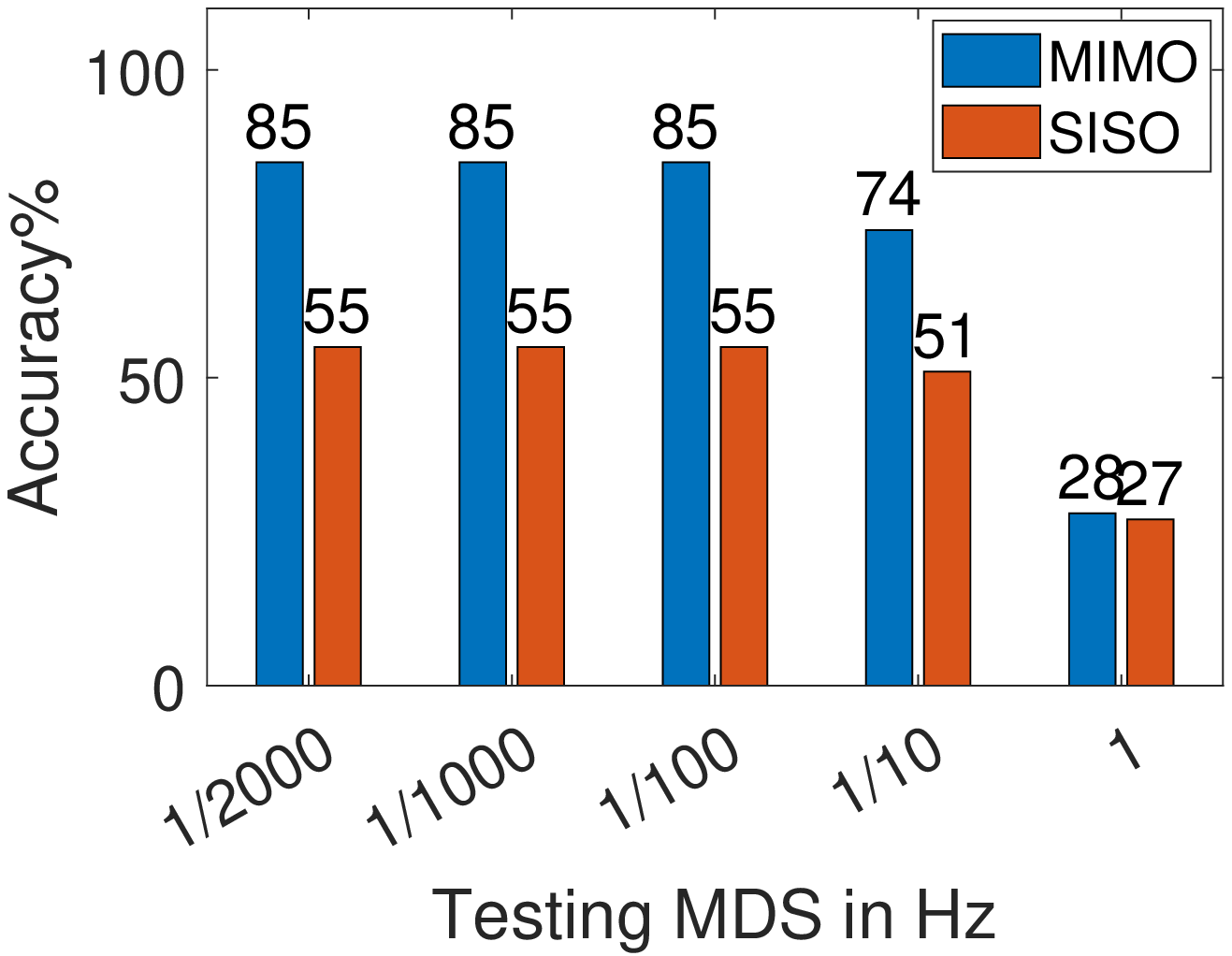}
    \label{subfig:fading7}}
    %
    \subfigure[Training MDS = 1/100 Hz ]
    {\includegraphics[width=0.4\columnwidth]{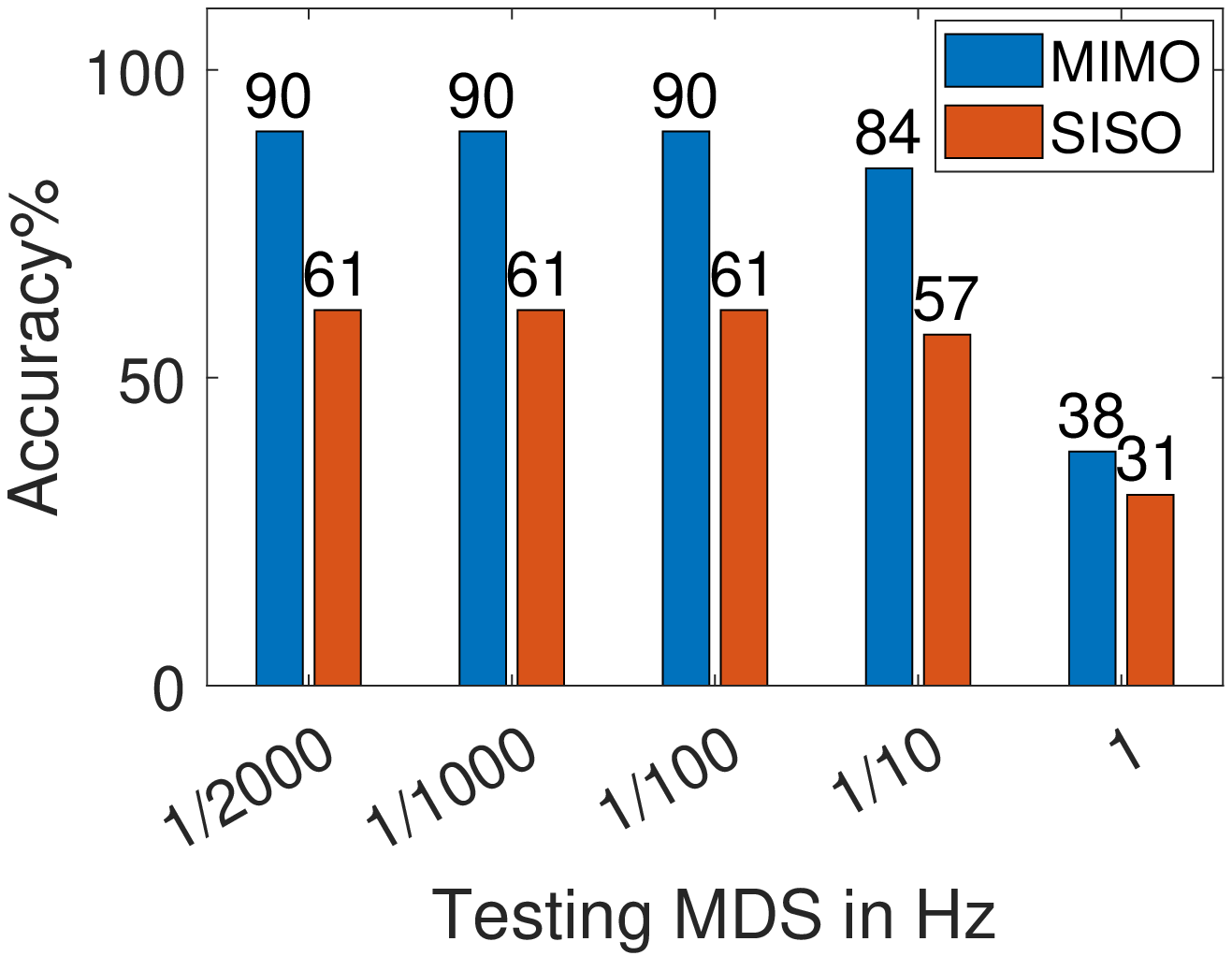}
    \label{subfig:fading8}}
    %
    \subfigure[Training MDS = 1/10 Hz]
    {\includegraphics[width=0.4\columnwidth]{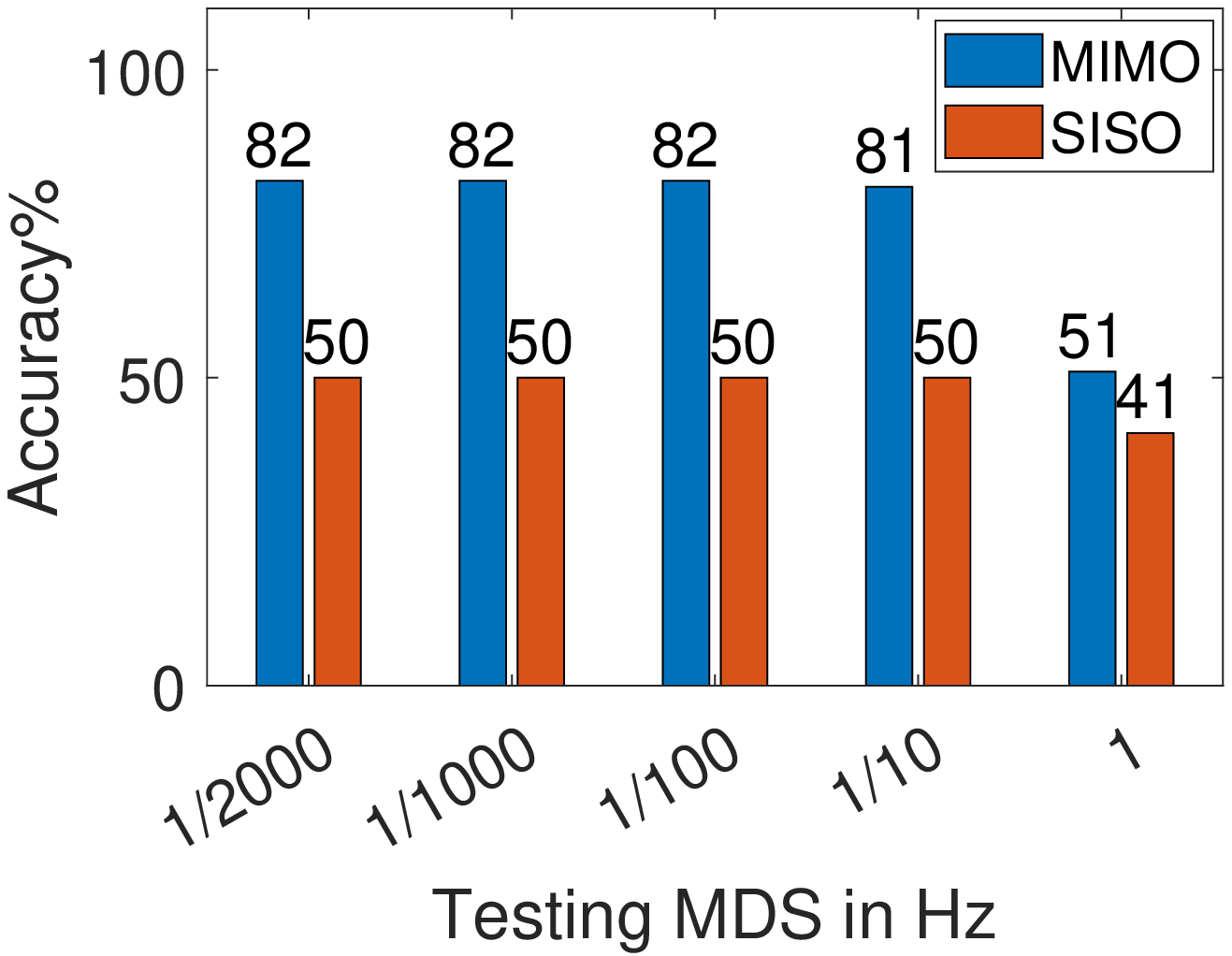}
    \label{subfig:fading9}}
    %
    \subfigure[Training MDS = 1 Hz ]
    {\includegraphics[width=0.4\columnwidth]{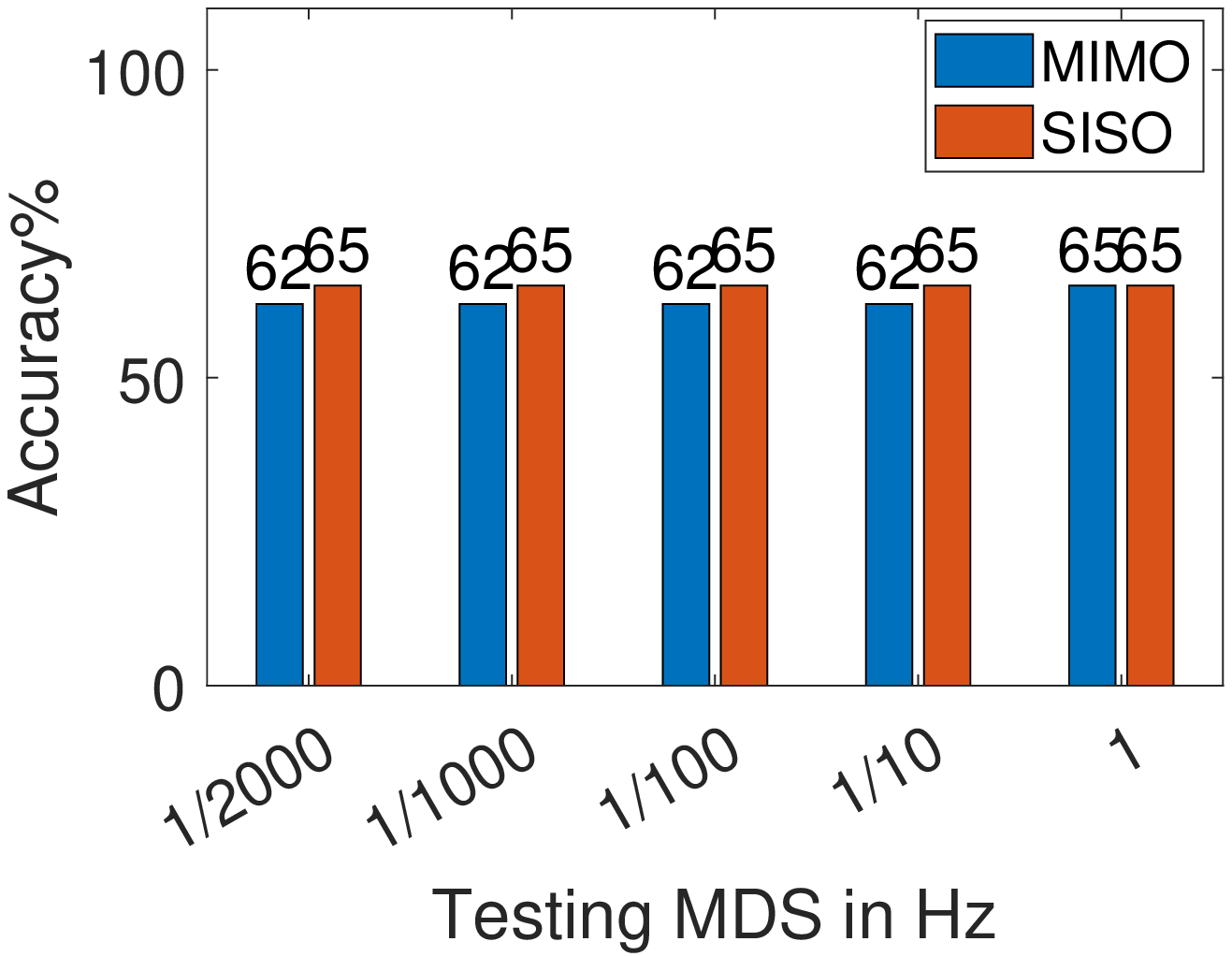}
    \label{subfig:fading10}}}
\caption{Impact of MDS on accuracy. APG is fixed at -20 dB.} 
\label{fig:diff training doppler shifts}
\end{figure*}

\paragraph{Impact of MDS: Testing Accuracy}
Figure~\ref{fig:diff training doppler shifts} shows the testing accuracy over Rayleigh channels where the CNN is trained and tested under various different MDS values, ranging from 1/2000 Hz (Figure~\ref{subfig:fading6}) to 1 Hz (Figure~\ref{subfig:fading10}), for both MIMO and SISO systems. The training APG value is fixed at -20 dB.
From this figure we observe that for small values of the training and testing MDS, the MIMO-based approach shows improvement in the testing accuracy over the conventional/SISO approach. However, at testing MDS of 1 Hz, the MIMO approach does not show any improvement over the SISO approach. We also observe that the MIMO approach is no longer agnostic to channel variation caused by the relative velocity between the transmitter and the receiver when the training and testing MDS values reaches 1 Hz. For instance, 1 Hz Doppler shift degrades the testing accuracy for the MIMO approach from about $90\%$ to $65\%$.
Looking at Figure~\ref{subfig:fading6}, which shows the testing accuracy under varied MDS values of the testing channel while fixing the training MDS to 1/2000 Hz, we observe that the MIMO approach achieves high testing accuracy for channels with testing MDS less than 1 Hz. When training MDS = 1/2000 Hz and testing MDS = 1/10 Hz, the MIMO-based approach achieves a testing accuracy of $83\%$ compared to $57\%$ for the conventional/SISO approach.    
We also observe that at testing MDS of 1 Hz, both MIMO and SISO approaches show degraded testing accuracy of about $30\%$. 
Figures~\ref{subfig:fading7} to~\ref{subfig:fading10}, depicting the testing accuracy for training MDS values ranging from 1/1000 to 1 Hz, show similar trends, except for Figure~\ref{subfig:fading10} corresponding to training MDS = 1 Hz, which shows that the testing accuracy under both approaches degrades to about $65\%$ with no improvement of MIMO over SISO. 
%

\begin{figure*}
\centerline{
    \subfigure[Training MDS = 1/2000 Hz]
   {\includegraphics[width=0.4\columnwidth]{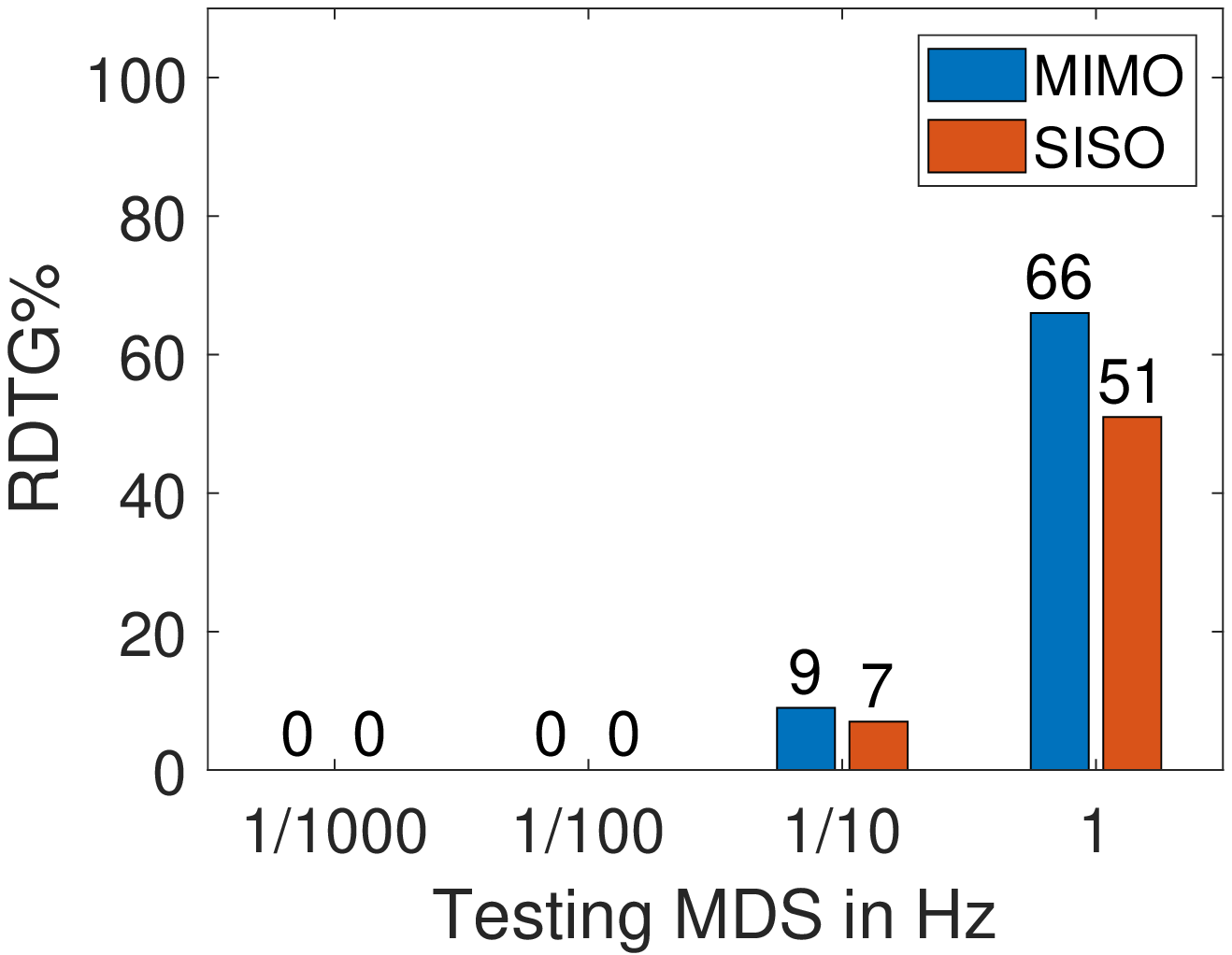}
    \label{subfig:RDTG-D1}}
    \subfigure[Training MDS = 1/1000 Hz]
   {\includegraphics[width=0.4\columnwidth]{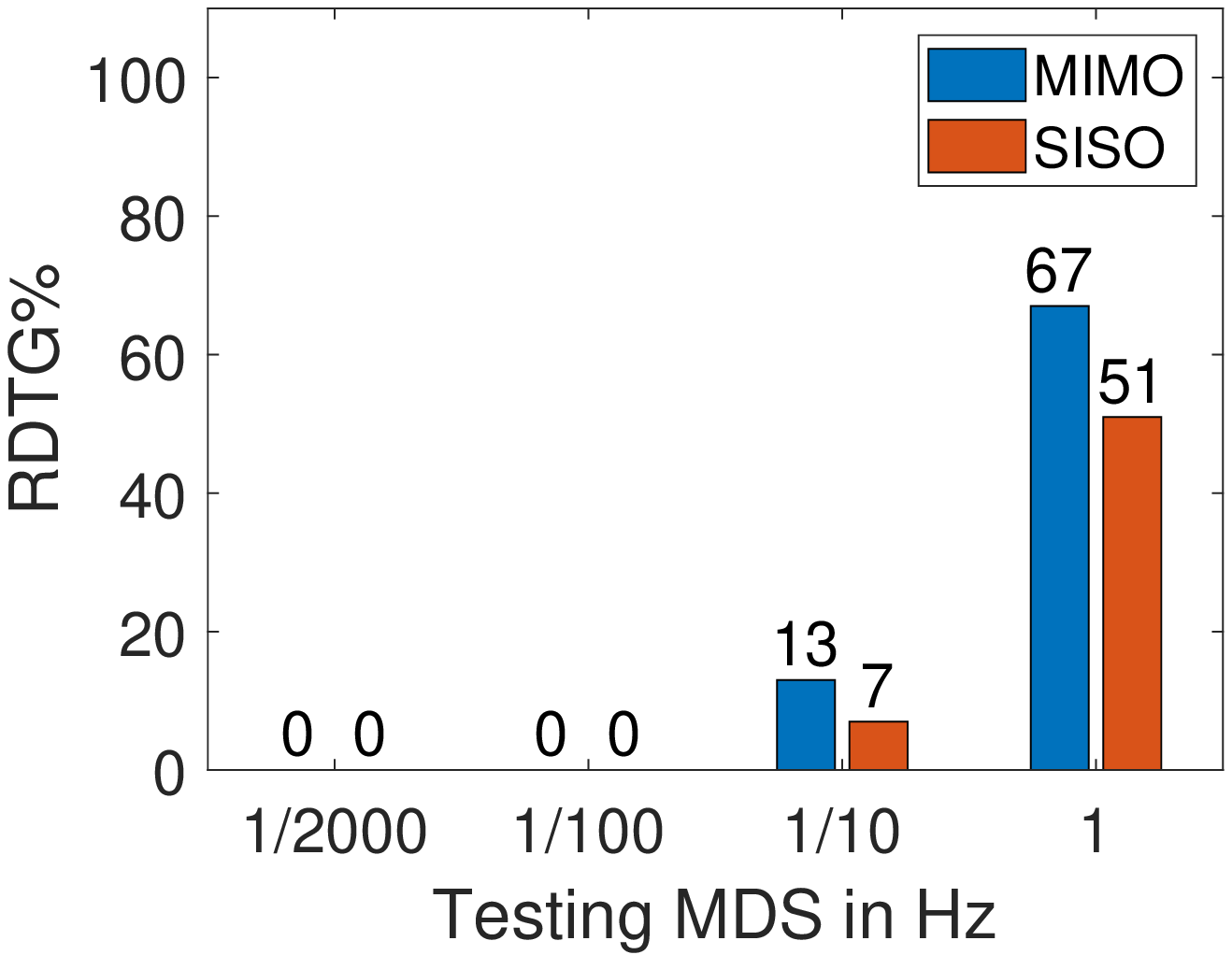}
    \label{subfig:RDTG-D2}}
    %
    \subfigure[Training MDS = 1/100 Hz ]
    {\includegraphics[width=0.4\columnwidth]{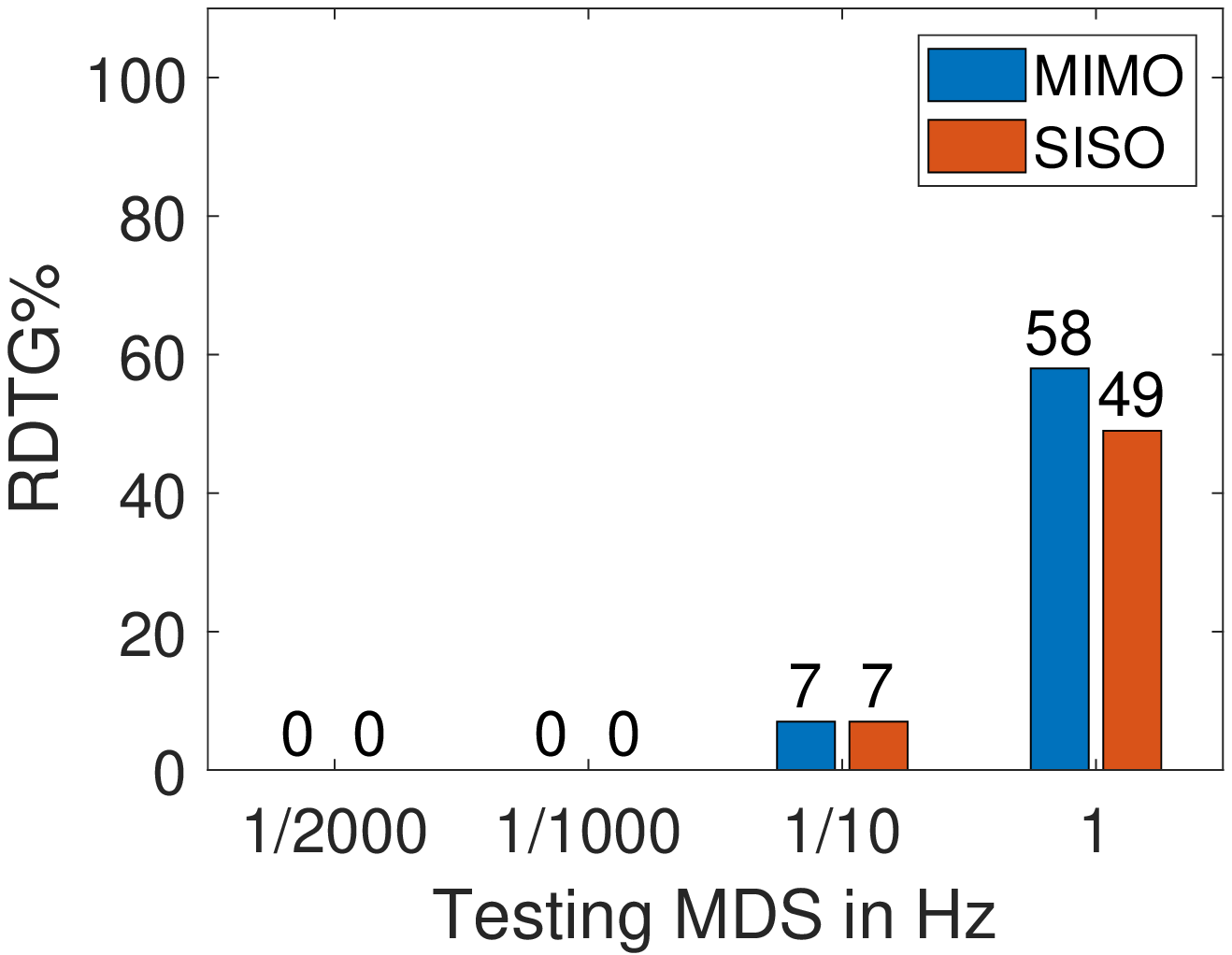}
    \label{subfig:RDTG-D3}}
    %
    \subfigure[Training MDS = 1/10 Hz]
    {\includegraphics[width=0.4\columnwidth]{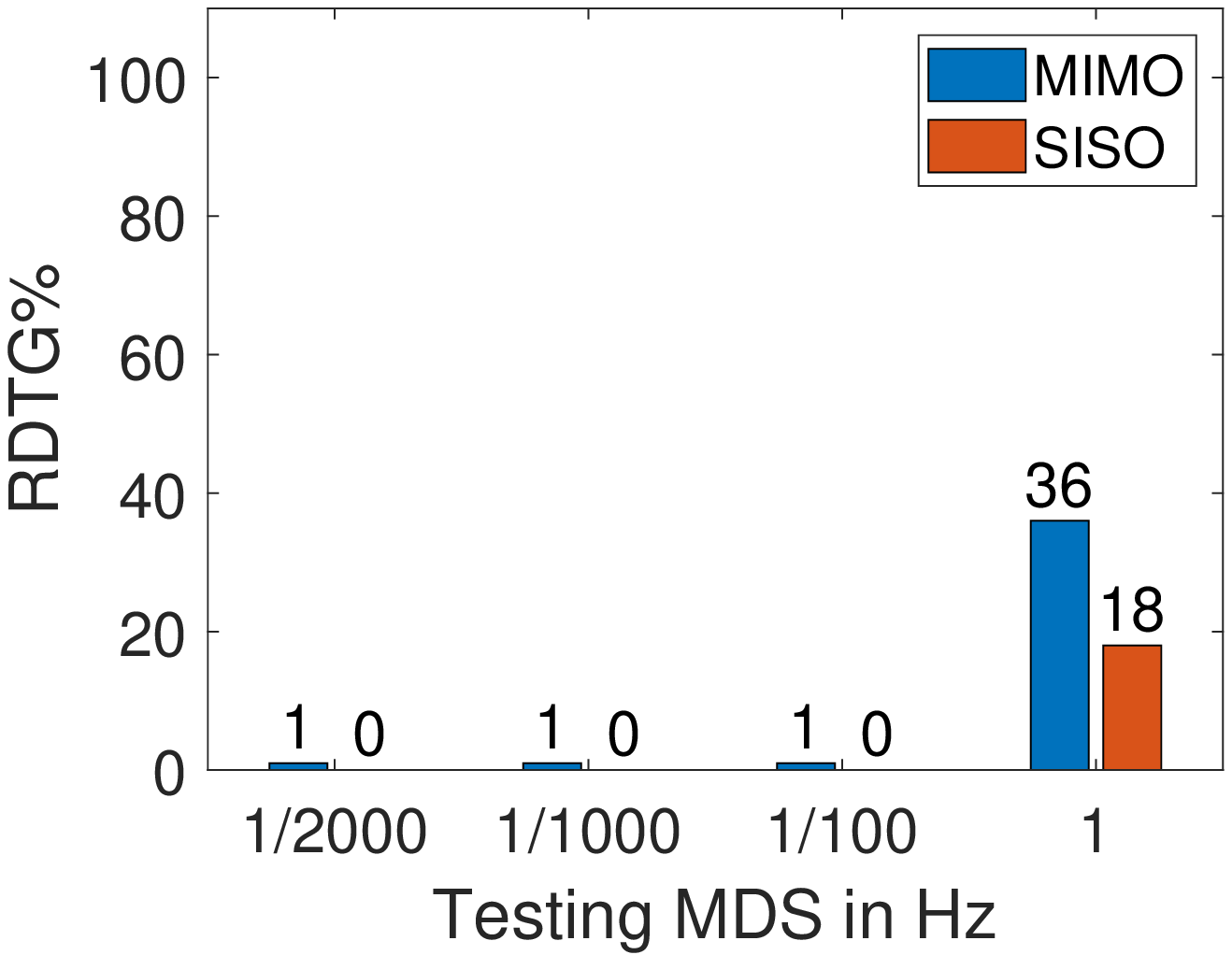}
    \label{subfig:RDTG-D4}}
    %
    \subfigure[Training MDS = 1 Hz ]
    {\includegraphics[width=0.4\columnwidth]{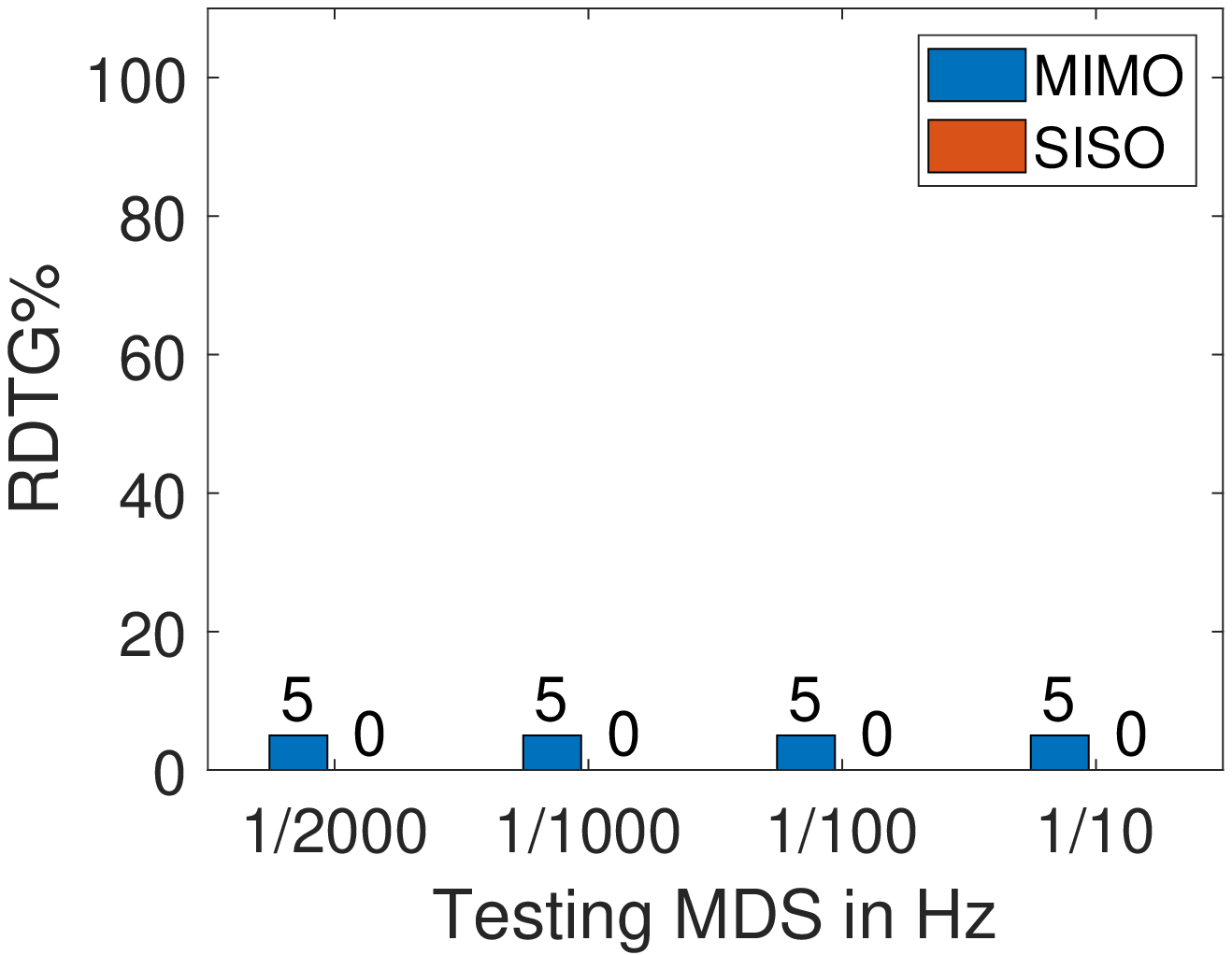}
    \label{subfig:RDTG-D5}}}
\caption{Impact of MDS on RDTG. APG is fixed at -20 dB.} 
\label{RDTG-MDS}
\end{figure*}

\paragraph{Impact of MDS: RDTG}
Figure~\ref{RDTG-MDS} shows the RDTG values where the CNN models are trained and tested under Rayleigh channels with different MDS values, ranging from 1/2000 Hz (Figure~\ref{subfig:RDTG-D1}) to 1 Hz (Figure~\ref{subfig:RDTG-D5}), and a training APG value fixed at -20 dB. 
This figure shows that for small testing MDS values, the testing accuracy achieved under both MIMO and SISO approaches is stable with an almost zero RDTG regardless of the training MDS. This is because small MDS values mean little to no variations in the channel conditions. 
However, when the testing MDS increases to 1 Hz, both MIMO and SISO approaches degrade in performance, with the MIMO approach achieving higher RDTG values, reflecting severe accuracy reduction and channel dependency.
These results show that the MIMO-based classification approach can mitigate flat slow fading channels. However, for fast fading channels, e.g. crowded indoor environments where the MDS values could exceed $30\ Hz$ at $3.6\ GHz$~\cite{hanssens_measurement-based_2016}, mitigating the wireless channel for reliable RF authentication remains an open challenge.

\section{Conclusion}
Although RF/device fingerprinting provides a lightweight technique proven resilient to spoofing that allows to identify illegitimate devices, its performance severely degrades by wireless channel condition variations, when the training channel differs from the testing channel. 
In this paper, we proposed a deep learning-based MIMO-enabled RF/device classification approach, and showed that the MIMO hardware capabilities can indeed mitigate the wireless channel effect and improve the RF fingerprinting accuracy for AWGN and flat fading channels.
We showed that, for AWGN channel, averaging the multiple received signals at the receiving end of a SIMO system of $L$ receiving antennas increases the SNR by a factor of $L$, leading to an improved classification accuracy. For channels with low SNR values, we conclude that the SNR gain provided by the SIMO-based approach compensates the SNR reduction.
We also showed that for flat fading channels, the proposed MIMO-based approach improves the classification accuracy by up to $60\%$ compared to the conventional/SISO approach, and that the improvement the MIMO approach achieves over the conventional approach is more significant and stable when the model is tested on less severe fading channels compared to the channel used for training.
We conclude that the CNN models tend to learn and extract features from both channel and device impairments, which results in a degradation the classification accuracy. 
We also conclude that despite the significant improvement the MIMO approach achieves in classification accuracy, this approach is suitable for slow flat fading channels, and mitigating the effect of the relative speed between the transmitter and the receiver is still an open challenge.

\label{sec:conc}

\section*{Acknowledgment}
This work was supported in part by the US National Science Foundation under NSF award No. 1923884.

\bibliographystyle{IEEEtran}
\bibliography{references}
\end{document}